\documentclass{aa}
\usepackage{graphicx}
\usepackage{txfonts}
\usepackage{diagbox}
\usepackage{rotating}
\usepackage{tikz}
\usepackage{multirow}
\usepackage{wasysym}
\usepackage{adjustbox}
\usepackage{afterpage}
\usepackage{placeins}
\usepackage{xcolor}
\usepackage{caption}
\usepackage[strict]{changepage}
\usepackage{lscape}
\usepackage{pdflscape}
\usepackage{soul}

\usepackage{hyperref}
\hypersetup{colorlinks=True,citecolor=blue}

\def\m2s2{\hbox{\,m$^{2}$\,s$^{-2}$} } %m2.s -2
\def\kms{\hbox{\,km\,s$^{-1}$} }       %km.s -1
\def\ang{\text{\AA}}

\def\pstret{\textbf{\emph{par\_stretching}}}
\def\prmax{\textbf{\emph{par\_Rmax}}}
\def\prmin{\textbf{\emph{par\_R}}}
\def\pfwhm{\textbf{\emph{par\_fwhm}}}
\def\pclim{\textbf{\emph{count\_cut\_lim}}}
\def\pcout{\textbf{\emph{count\_out\_lim}}}
\def\preg{\textbf{\emph{par\_reg\_nu}}}
\def\psbox{\textbf{\emph{par\_smoothing\_box}}}
\def\pskernel{\textbf{\emph{par\_smoothing\_kernel}}}
\def\pvici{\textbf{\emph{par\_vicinity}}}
\def\pmask{\textbf{\emph{mask\_telluric}}}
\def\pccf{\textbf{\emph{CCF\_mask}}}
\def\denoisdist{\textbf{\emph{denoising\_dist}}}

\def\pfeedback{\textbf{\emph{feedback}}}

\usepackage{amstext}

\begin{document}

   \title{RASSINE: Interactive tool for normalising stellar spectra}

   \subtitle{I. Description and performance of the code}

   \author{M. Cretignier\inst{1}
          \and J. Francfort \inst{2}
          \and X. Dumusque \inst{1}
          \and R. Allart \inst{1}
          \and F. Pepe \inst{1}
          }

   \institute{Astronomy Department of the University of Geneva, 51 ch. des Maillettes, 1290 Versoix, Switzerland\\
              \email{michael.cretignier@unige.ch}
         \and
                      D\'epartement de Physique Th\'eorique, Universit\'e de Gen\`eve, 24 quai Ansermet, CH–1211 Gen\`eve 4, Switzerland\\
             \email{jeremie.francfort@unige.ch}
             }

   \date{Received XXX ; accepted XXX}

% \abstract{}{}{}{}{} 
% 5 {} token are mandatory
 
  \abstract
  % context heading (optional)
  % {} leave it empty if necessary  
   {}
  % aims heading (mandatory)
   {We provide an open-source code allowing an easy, intuitive, and robust normalisation of spectra.}
  % methods heading (mandatory)
   {We developed RASSINE, a \emph{Python} code for normalising merged \emph{\textup{1D}} spectra through the concepts of convex hulls. The code uses six parameters that can be easily fine-tuned. The code also provides a complete user-friendly interactive interface, including graphical feedback, that helps the user to choose the parameters as easily as possible. To facilitate the normalisation even further, RASSINE can provide a first guess for the parameters that are derived directly from the merged \emph{\textup{1D}} spectrum based on previously performed calibrations.}
  % results heading (mandatory)
   {For HARPS spectra of the Sun that were obtained with the HELIOS solar telescope, a continuum accuracy of $0.20\%$ on line depth can be reached after normalisation with RASSINE. This is three times better than with the commonly used method of polynomial fitting. For HARPS spectra of $\alpha$ Cen B, a continuum accuracy of $2.0\%$ is reached. This rather poor accuracy is mainly due to molecular band absorption and the high density of spectral lines in the bluest part of the  merged \emph{\textup{1D}} spectrum. When wavelengths shorter than 4500 \ang{} are excluded, the continuum accuracy improves by up to $1.2\%$.  The line-depth precision on individual spectrum normalisation is estimated to be $\sim 0.15\,\%$, which can be reduced to the photon-noise limit ($0.10\,\%$) when a time series of spectra is given as input for RASSINE.}  
   {With a continuum accuracy higher than the polynomial fitting method and a line-depth precision compatible with photon noise, RASSINE is a tool that can find applications in numerous cases, for example stellar parameter determination, transmission spectroscopy of exoplanet atmospheres, or activity-sensitive line detection.}


   \maketitle

\section{Introduction}

A spectrum is a fundamental observable that is used to study astronomical objects such as galaxies, stars, and exoplanets. It describes the distribution of photons per wavelength bins and can either be used in terms of absolute quantity to determine the luminosity of the objects, or colours can be analysed separately through photometric bands. A rich content of information is also brought by the absorption or emission lines, for which the spectrum has to be normalised by a continuum. This happens for instance in the framework of stellar abundances studies \citep[e.g.][]{Blanco(2014),Sousa(2015),Adibekyan(2016)} or exoplanet atmospheres \citep[e.g.][]{Wyttenbach(2015),Allart(2017)}. For radial velocity (RV), the spectrum does not need to be continuum-normalised, but a colour correction has to be applied \citep[e.g.][]{Bourrier(2014),Malavolta(2017)}, which is itself related to the continuum of the spectrum. Normalised spectra are also necessary to construct the binary masks used in Doppler spectroscopy to extract RV using the cross-correlation technique \citep{Pepe(2003)}. Finally, a precise determination of the continuum level also allows observing the stellar line variability that is induced by the stellar activity \citep{Thompson(2017),Wise(2018), Dumusque(2018), Cretignier(2019)}.%, the exception is made of spectral type higher than F for which the spectra are directly analysed with the blaze function \citep{Becker(2015)}. 

When absorption lines are studied, a non-trivial step consists of normalising the spectrum by its continuum, where the latter can differ from a black-body curve due to Rayleigh scattering reddening. This differential chromatic response can also be directly induced by the spectrograph itself through the optical elements and the CCD quantum efficiency, which varies with wavelength. An effective
process for normalising high-resolution spectra obtained by various surveys in a unified and coherent way appears as an important
step. These surveys include ESPRESSO \citep{Pepe(2014)}, EXPRES \citep{Fischer(2017)}, NEID \citep{Schwab(2016)}, PEPSI \citep{Strassmeier(2015)}, CRIRES, NIRPS \citep{Bouchy(2017)}, and previous surveys, such as CORALIE \citep{Queloz(2000)}, HARPS \citep{Pepe(2002b), Mayor(2003), Pepe(2003)}, HIRES \citep{Pasquini(2010)}, or HARPS-N \citep{Cosentino(2012)}, .
%a code to normalise them in the most coherent way appears as a relevant step.\commick{reformuler} \\

Current methods often deal with the individual orders of \emph{\textup{1D}} extracted echelle-order spectra because they represent a narrower band of the spectrum where the continuum presents fewer inflection points. A filter is often used in order to smooth the data and remove as many stellar lines as possible. This filter can be a rolling maximum or moving average, an asymmetric sigma clipping, or even a Fourier filtering. The continuum is then estimated by fitting a low-order polynomial on the filtered spectrum \citep{Tody(1986),Tody(1993)}.

When the orders are blaze-corrected and merged to produce a single merged \emph{\textup{1D}} spectrum with a large wavelength coverage, a model with many free parameters is necessary to account for the numerous inflection points in the continuum, which are induced by atmospheric absorption and CCD response. A high-order polynomial is often not flexible enough. When a stellar template is available for the star or spectral type for which we wish to normalise the spectrum, it is possible to calculate the ratio between the spectrum we wish to normalise and the template, and fit any trend observed in this ratio. This trend is then removed from the spectrum to normalise it. The main advantage of this method consists of considerably reducing the required order of the polynomial, but the intrinsic disadvantages of polynomial fitting remain \citep{Skoda(2008)}. A method like this allows for instance to correct for colour variation induced by different airmass observations \citep{Malavolta(2017)}.

The rolling alpha shape for a spectrally improved normalisation estimation (RASSINE) is an open-source\footnote{\url{https://github.com/MichaelCretignier/Rassine\_public}} Python code \citep{Rossum(1995),Rossum(2009),Python(2019)} for normalising merged \emph{\textup{1D}} spectra, which are obtained by merging blaze-corrected \emph{\textup{1D}} echelle-order spectra. The code provides interactive graphical user interfaces (GUI) to help users through the different steps that are required to efficiently normalise a spectrum. The code uses standard Python libraries such as Numpy \citep{Numpy(2006),Walt(2011)}, Matplotlib \citep{Matplotlib(2007)}, Scipy \citep{Scipy(2020)}, Pandas \citep{Pandas(2010)}, and Astropy \citep{Astropy(2018)}. 

RASSINE uses the convex-hull and alpha-shape theories to model the upper envelope of a spectrum, which is often equivalent to the stellar continuum for solar-type stars. We note that the algorithm used by RASSINE to normalise spectra is similar to the AFS code that was recently published in \citet{Xu(2019)}, where the authors showed the higher performance of the alpha-shape method compared to classical iterative methods. RASSINE and AFS are different, however, because the former uses merged \emph{\textup{1D}} spectra as input while the latter uses \emph{\textup{1D}} echelle-order extracted spectra. We note that RASSINE uses six free parameters to model the spectrum continuum. An automatic mode of the algorithm, described in Appendix \ref{sec:auto}, can provide first guesses for five of them.

The theory behind RASSINE is described in Sect.~\ref{sec:theory}. The continuum accuracy and line-depth precision of the code on spectral line depth, as well as its ability to extract the correct continuum around broad absorption lines, are described in Sect.~\ref{sec:perf}. We then conclude in Sect.~\ref{Conclusion}.

\section{Theory}\label{sec:theory}

\subsection{Convex hull and alpha shape}
Determining the upper envelope of a spectrum is closely related to the concepts of convex hull and alpha shape, the latter being a generalisation of the former. We briefly review these ideas here and restrict our description to the 2D case.

The convex hull of a set of points $S$ in the plane can be understood instinctively as follows: take an elastic band and stretch it around the set $S$. The final shape that the band will take is the convex hull of the set. Mathematically speaking, if the set is made of points $x_i$, the convex hull $C(S)$ can be written as \citep{Asaeedi(2013)}  
\begin{equation}
    C(S) = \left\{y_i = \sum \alpha_i x_i \in \mathbb{R}^2\ \biggr\vert\ \alpha_i \geq 0, \sum \alpha_i =1 \right\}.
\end{equation}

The convexity of the convex hull forbids the existence of inflexion points. Because a black-body radiation curve presents inflexion points, we need to consider the concept of alpha shapes, which is a generalisation of the convex-hull theory and allows for concave hulls.\

When a convex hull is considered, all the internal angles are at most $180^\circ$. In the alpha-shape framework, this condition is relaxed such that each internal angle is at most $180^\circ + \alpha$. This allows modelling the inflexion points in the upper envelope of a spectrum. This upper envelope is then modelled by a shape that is intermediate between a convex hull ($\alpha=0^\circ$) and a full alpha shape ($\alpha = 180^\circ$).

\subsection{Outline and structure of the code \label{sec:code}}

\subsubsection{Brief overview}

The main assumption behind the code is that the local maxima of the spectrum correspond to the continuum. This assumption is relatively well satisfied for solar-type spectra, except perhaps in the bluest part of the visible because of  atmosphere extinction and the high density of spectral lines. However, this assumption no longer holds for M-dwarf spectra, for which the extremely high density of lines completely erases the stellar continuum at nearly all wavelengths.

RASSINE can be subdivided into five steps known as the SNAKE sequence:
\begin{enumerate}
    \item Smoothing of the spectrum,
    \item neighborhood maxima detection,
    \item alpha-shape algorithm,
    \item killing outliers,
    \item envelope interpolation.
\end{enumerate}
The smoothing is necessary to increase the signal-to-noise ratio (S/N) per bin element, which allows handling spectra at different noise levels. This step is also relevant to prevent adjusting the continuum to the upper envelope of the noise in spectra with low S/N. After the smoothing, all the local maxima are first flagged. The maxima corresponding to the upper envelope are then chosen using a modified gift-wrapping algorithm, which removes the local maxima formed by blended lines. This alpha-shape algorithm can be compared to a rolling pin that rolls over the local maxima.  
%After smoothing, the maxima corresponding to the upper envelope are chosen using a modified gift-wrapping algorithm, which reject local maxima induced by blended lines. An alpha-shape, that can be compared to a "rolling pin" rolling over the remaining local maxima, is then used to find the {\bf local maxima corresponding to the stellar continuum. 
The code finally rejects remaining outliers that exist in the final selection of local maxima and performs different types of interpolation on the remaining maxima to obtain the stellar continuum.

\begin{figure*}[h]
        \includegraphics[width=18cm]{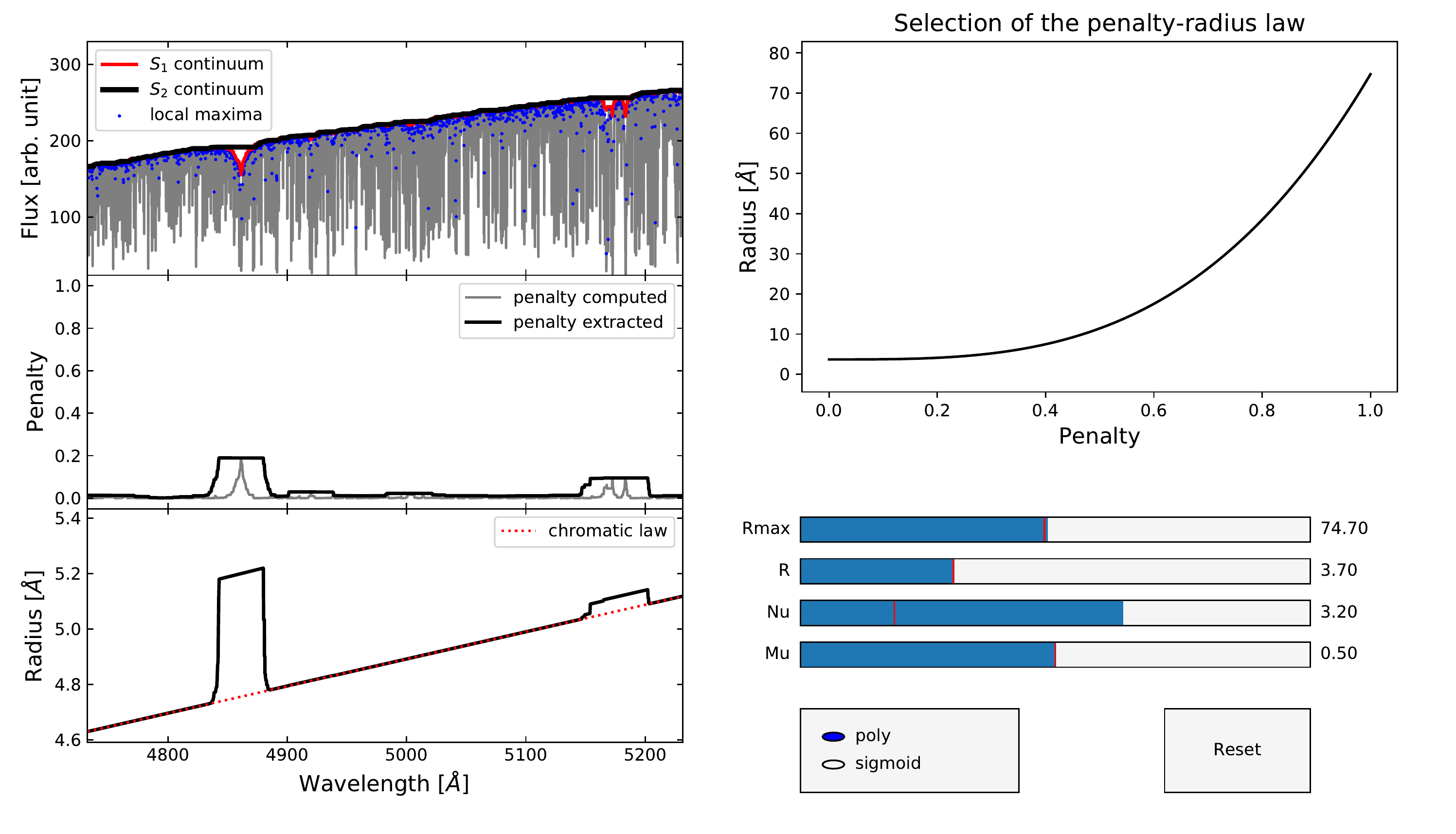}
        \caption{ Second GUI of RASSINE\ that is used to define the varying radius of the alpha shape. In this case, the values for the parameters are bad ($\nu$ too high), but they are chosen to highlight the dependence of the radius on wavelength. \textbf{Top left:} Local maxima (blue dots) extracted from the smoothed spectrum (grey curve). A continuum $S_1$ using a small window (red curve) and $S_2$ using a large window (black curve) are computed with a rolling maximum on the spectrum. \textbf{Middle left:} Penalty computed from the relative difference between $S_1$ and $S_2$ (grey curve). The penalty is rounded and looks like a step function (black curve). \textbf{Bottom left:} Final radius values for the rolling pin along the wavelength axis. Broader absorption lines such as H$_{\beta}$ at 4861 $\ang$ or the MgI triplet around 5172 $\ang$ are penalized such that the rolling pin increases in size. The observed linear trend corresponds to the dispersion because all spectral lines at first order have the same line width in velocity space. \textbf{Top right:} Penalty map relation between the radius of the rolling pin and the penalty value. The curve is parametrised with a minimum radius $R$, a maximum radius $R_{max}$ , and an exponent $\nu$ that represents the exponent of the power law. The second parameter $\mu$ is only useful when the sigmoid function is selected for the penalty map and represents the centre of the sigmoid ($\nu$ being proportional to the sigmoid width).}
        \label{FigInterface1}
\end{figure*}

\subsubsection{Smoothing of the spectrum\label{sec:smooth}}

This first step of smoothing consists of removing the high-frequency noise from the stellar continuum to prevent contamination by it when the best fit to the continuum is determined. RASSINE proposes two different smoothing options, either degrading the spectrum resolution, or using a Savitzky-Golay filter \citep{Savitzky(1964)}.

For the first smoothing option, the spectral resolution is degraded by convolving the spectrum with either a rectangular or Gaussian kernel. We note that this convolution will change the depth of the spectral line, but because the goal is to best fit the stellar continuum, this is not a problem. The kernels can be chosen with the parameter \pskernel{}. We use the convention of the bold and italic font for the parameters from here on, where the name of the parameter is written as in the config file of the code. The strength of the smoothing is controlled by the width of the kernel, which can be tuned by the \psbox{} parameter (in units of wavelength sampling).

 The second smoothing option consists of a Savitzky-Golay filter, which can be described as a low-order polynomial rolling fit on the data. The polynomial is chosen to be of the third degree. The advantage of this filter is its ability to keep the global shape of the spectrum almost invariant, whereas it filters the oscillations on a scale smaller than the window very well. However, this filter remains more sensitive to strong flux variation, which can be due to cosmic rays, lamp contamination, or hot pixels, for example. To counteract this problem, an asymmetric sigma-clipping is performed on the absolute difference between the smoothed and initial spectrum, and outliers in the smoothed spectrum are brought back to their initial flux values to detect and reject these points more easily at a later stage.

\subsubsection{Local maxima}
The second step performed by the code is to search for the local maxima.
A point is called a local maximum if its value is the highest in its closest neighbourhood, which can be tuned by the parameter \pvici{} that corresponds to the size of the half-window defining the neighbourhood (in units of the wavelength sampling, e.g. 0.01 \AA\,for HARPS spectra). If the previous smoothing step was performed correctly, this parameter remains quite irrelevant, but it becomes important when low S/N spectra are studied to speed up the code.

\subsubsection{Penalty for broad absorption regions \label{sec:penal}}

As we show in the next sub-section, the idea behind RASSINE is to roll a rolling pin, of a size larger than typical spectral lines, on top of the spectrum to adjust its continuum. At first order, all the spectral lines should have the same width in velocity space, determined by the projected stellar rotation, macroturbulence, and instrumental resolution \citep{Gray(2005)}. It is therefore important to determine an optimal value for the average line width, which is stored in the \pfwhm{} variable (units of \kms). This can be done by using the full width at half-maximum (FWHM) of a Gaussian fitted to the cross-correlation function (CCF, \citet{Baranne(1996),Pepe(2002)}).

However, if the size of the rolling pin is always the same, the continuum will be poorly fitted in region of strong absorption. When lines begin to saturate, for example for the H$_\alpha$, CaII H\&K, sodium doublet NaD or triplet magnesium MgIb absorption lines, the wings of the Voigt profile begin to dominate and the line width diverges from the typical value of the weak regime.  To prevent the rolling pin from falling in regions of broad absorption, a penalty map is used to increase the radius of the rolling pin as soon as such a region is reached. The GUI for this step is presented in Fig.~\ref{FigInterface1}. First of all, we adjust two approximated continua of the spectrum using a rolling maximum algorithm. The first, $S_1$ (red curve in the top left panel of Fig.~\ref{FigInterface1}), using a small window of 40 times the FWHM of the CCF (converted into \ang{}  using $\lambda_{min}$, the minimum wavelength of the analysed spectrum), and the second, $S_2$, using a window $10$ times larger than $S_1$ (black curve in the top left panel of Fig.~\ref{FigInterface1}). Even when both windows are fixed in wavelength, it is not necessary to increase them in size with the dispersion because both of them are already far more extended than the typical line width. To prevent the rolling maximum algorithm to be sensitive to anomalous flux intensities (cosmic rays or contamination between fibers), we first reject outliers using a rolling sigma-clipping on a 5 \AA\,window, with the specificity that in our sigma-clipping, the mean is replaced by the median and the standard deviation by 1.5 time the interquartile range (1.5 time the 75\% - 25\% quartiles,  \citep{Upton(1996)}). Then, for each wavelength, the penalty $p=(S_2-S_1)/S_2$ is computed and normalised by the minimum and maximum value over the entire spectrum to lie between $0$ and $1$. This penalty curve (grey curve in the centre left panel of Fig.~\ref{FigInterface1}) is then transformed into a step function (black curve in the same panel) to make the code more efficient.

The next step consists of choosing the parameter \preg{}, which appears in the function used to map the penalty $p$ to the radius $r$ of the rolling pin (black curve in the bottom left panel of Fig.~\ref{FigInterface1}). Two functions to map the penalty to the radius of the rolling pin are provided: a polynomial function, and a sigmoid one. For polynomial mapping (black curve in the top right panel of Fig.~\ref{FigInterface1}), the radius is given by 
\begin{equation}
    r(p,\lambda) = C(\lambda)[R + (R_{max} - R)\cdot p^\nu]
,\end{equation}
where $C(\lambda)$ is the chromatic law. As
line widths are similar at first order in velocity space, the classical Doppler effect formula tells us that line width in wavelength-space will be a linear function of wavelength. Therefore the radius of the rolling pin in wavelength should satisfy the chromatic law $C(\lambda) = \lambda /\lambda_{min}$. $\nu$ is a real positive parameter specified directly with \preg{}, and $R$, $R_{max}$ are two parameters of the model: \prmin{} and \prmax{}. They define the minimum and maximum radii in wavelength units (generally \AA). 
Both parameters must be specified at the minimum wavelength of the spectrum because the values will be scaled by the C($\lambda$) in the penalty law. The values for \prmin{} and \prmax{} have to be provided by the user regarding the typical line width of its spectrum as well as the broadest absorption gap, but both parameters can be estimated using the automatic mode (see Appendix \ref{sec:auto}). If $\nu>1$, the penalty map is convex, and only high penalties will modify the radius substantially. In contrast, if $\nu<1$, the map is concave, and only low penalties will leave the radius unchanged. The second available law, the sigmoid (see Fig.~\ref{FigInt2} in Appendix \ref{app:interface} for an example), allows us to produce a two-radius regime because\ its shape is step-like. The transition from the smallest and largest radius is given by the sigmoid centre $\mu,$ and the smoothness of the transition is given by the sigmoid width $\nu$. The best value for a given spectrum depends on its shape. Smoothed spectra can be effectively reduced with $\nu<1,$ whereas spectra with long oscillations due to instrumental effects or poor merging of \emph{\textup{1D}} extracted echelle-order spectra are better reduced with $\nu>1$.  A good value for several cases was found using the polynomial mapping and $0.7<\nu<1.3$. This parameter is the only one for which no calibration was performed to adjust it automatically.

\begin{figure}[tp]
        \centering
        \includegraphics[width=9cm]{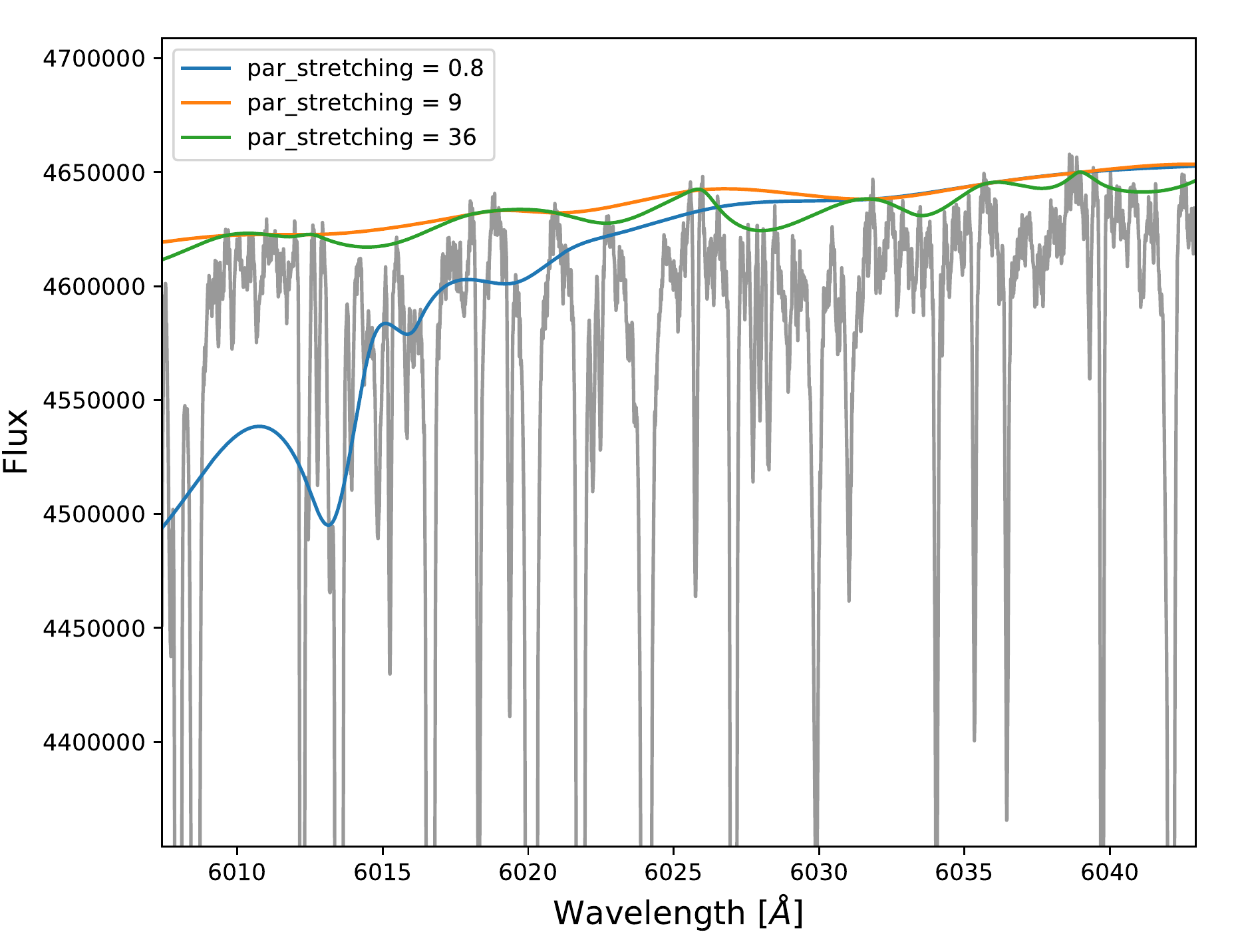}
        \caption{Effect of different values for \pstret{} on the normalisation of the spectrum (grey curve). When \pstret{} is too small (blue curve), the alpha shape falls on the blended local maxima. As soon as \pstret{} is sufficiently high (orange curve), the alpha-shape approach provides a good normalisation. %If the continuum is compared to a veil, the parameter plays the role of a tension and a too high value (green curve) reduces the tension. 
        A wide range of values (orange and green curves) can provide a good normalisation below the 1\% level. %According to our calibrations, \pstret{} always has to be higher than 2.
        }
        \label{FigStretch}
\end{figure}

\subsubsection{Alpha shape (rolling pin) \label{sec:alpha}}
This part is the main step of RASSINE. At this stage, we have a collection of selected local maxima from which we would like to extract a continuum. Moreover, for each of these local maxima, a radius for the alpha shape was assigned in the previous section (Sect.~\ref{sec:penal}). The idea is the following: the alpha shape, which can be seen as a rolling pin, starts on the maxima with the smallest wavelength. It then rolls clockwise, being anchored to this maximum, until it touches another maximum. This maximum is then taken to be the next anchor point, the radius is updated, and the code continues iteratively. An algorithm like this is called gift wrapping or Jarvis walk \citep{Jarvis(1972)} and is presented in appendix \ref{sec:as_algo}. Other algorithms for a convex hull exist \citep[e.g.][]{Graham(1972),Eddy(1977),Phan(2007)}, but the Jarvis walk remains the most intuitive.

The alpha shape, or more intuitively, the rolling pin, is a circle that rolls on top of the spectrum. However, when we take the raw spectrum, we have a problem because flux units, normally expressed in analog-to-digital units (ADU), are much larger than the wavelength units, which are generally expressed in $\AA$. We therefore have to renormalise the flux so that variations in the wavelength and flux directions are comparable. This is done by transforming the flux according to the following relation: $f = f_{raw} \frac{\Delta \lambda}{\Delta f_{raw}}$, with $f_{raw}$ the flux of the raw spectrum, and $\Delta \lambda$ and $\Delta f_{raw}$ the difference between the maximum and minimum wavelength and raw flux values, respectively. After this first scaling, the following equality is satisfied: $\lambda_{max} - \lambda_{min} = f_{max} - f_{min}$. A second stretching is then applied using the parameter \pstret{} , which is positive and larger than $1$, which leads to the following relation:
\begin{equation}
\lambda_{max} - \lambda_{min} = \pstret{} \cdot (f_{max} - f_{min}).
\end{equation} 
  
The scaling of the axes is critical for the success of the alpha-shape step performed later. An inappropriate value can lead to unsuitable normalisation where the continuum is not on top of the spectrum, but is rather going through it (see the blue curve in Fig.~\ref{FigStretch}). This can be explained by the fact that the alpha shape rolls on the spectrum from the left to the right and that the algorithm is only applied to local maxima. If the stretching parameter is too small, the good local maxima are beyond reach and the alpha shape chooses local maxima created by blended lines to continue rolling. When the stretching is well selected, the alpha shape rolls on top of the spectrum, thereby selecting local maxima corresponding to the continuum. Details about this process and the selection of local maxima are described in Appendix \ref{sec:as_algo}. Hopefully, as was concluded in \citet{Xu(2019)}, a wide range of values for the stretching produces good enough results and incorrect values are easily recognisable in the final product. For intuition, the \pstret{} parameter can be seen as the inverse tension of a veil that would cover the spectrum; the lower its value, the higher the tension.

\begin{figure*}[h]
        \includegraphics[width=18.5cm]{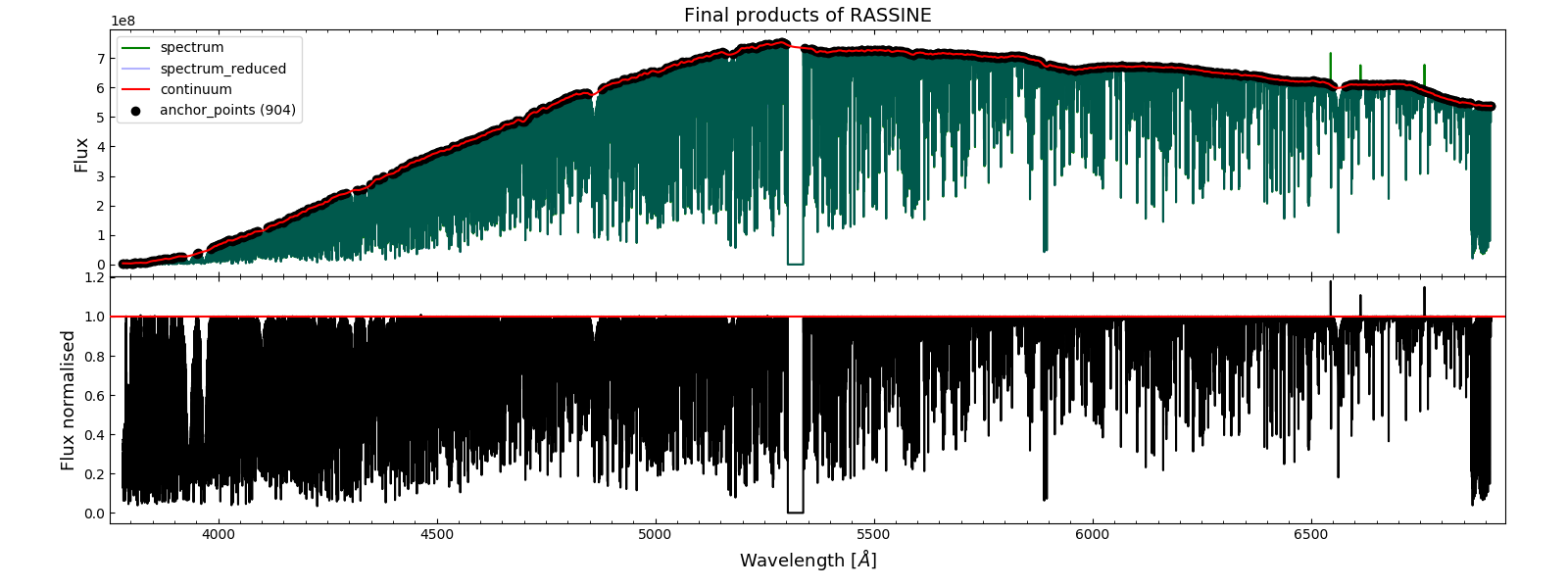}
        \caption{Final plot of the GUI. \textbf{Top:} Initial solar spectrum given as input to RASSINE (green curve), as well as the continuum (red curve) as computed by RASSINE in complete automatic mode (see Appendix \ref{sec:auto}). This continuum is derived using a cubic interpolation over a selection of relevant local maxima that are also called anchor points (black dots). The continuum avoids three spurious peaks due to simulated cosmic rays at $6550$, $6600,$ and $6750\,\AA$ and jumps above the large absorption regions. \textbf{Bottom:} Spectrum normalised with RASSINE.}
        \label{FigInterface2}
\end{figure*}

\subsubsection{Outlier detection\label{sec:outliers}}

Algorithms for selecting an upper envelope remain very sensitive to outliers with anomalous flux intensities. Even though much attention has been expended so far into suppressing as many of these points as possible, we perform here a final check by considering three types of outliers: 1) edge maxima, 2) high or low extrema, and 3) maxima that are too close. 

\begin{enumerate}
    \item{At the blue and red edges of the spectrum, the rolling pin touches spurious maxima simply because it starts and ends on them. If the spectrum were measured on an infinite wavelength grid, the continuum would touch other maxima, which are unfortunately not available. To prevent selecting those spurious maxima as part of the continuum, we associate with the $\pclim{}$ first and $\pclim{}$ last maxima the value of the $\pclim{}+1$ and $N-\pclim{}-1$ maxima, where $N$ is the number of selected maxima.
By default, $\pclim{}$ is set to three, therefore RASSINE flattens the first and last three maxima. This step can be performed with the GUI (see Fig.~\ref{FigInt3} in the appendix).}

\item{
Sharp peaks are removed by computing the left and right derivative at each selected maximum with respect to the two neighbouring maxima. By default, the five points with the highest difference between the left and right absolute derivatives are displayed by the GUI, and the user can choose to retain or delete them (see Fig.~\ref{FigInt4} in the appendix). If one point is suppressed, the next one with the highest absolute derivative difference will replace it. By default, RASSINE suppresses the points that are higher than the 99.5 percentiles of the distribution. The process can be run iteratively, with the number of iterations controlled by the optional parameter \pcout.} 

\item{
Maxima that are too close to each other are problematic for the cubic interpolation used at the end to define the continuum (see Sect.~\ref{sec:env}). Moreover, because these points do not add any relevant information, they are removed automatically. To detect them, the code computes the distance between each pair of neighbouring maxima that is still present in the final selection. Then, an asymmetric sigma-clipping is performed on the distribution of the distance difference to identify outliers. For each group of detected close maxima, the algorithm retains the point that maximises the equidistance with the two neighbouring maxima.}

\end{enumerate}

%for the $j^{th}$ maxima, RASSINE computes the difference of the distance between the $(j+1)^{th}$ one and $(j-1)^{th}$ one, namely the difference of the left and right closest neighborhood distance. An asymmetric sigma clipping is performed on the distribution of the distance difference to identify the doubtful points. The algorithm then removes between the $(j-1)^{th}$, $j^{th}$ and $(j+1)^{th}$  point the one for which the suppression produce the most equidistant residual grid.

%MODE D'EMPLOI If all these automatic criteria are not sufficient, the user can, in the last interactive window, specify by hand the index of the points to delete. In this window, the local maxima are numbered such that the user just has to enter the list of indices in the console. The new continuum is then displayed and the user has to valid or cancel their choice depending on their satisfaction of the new continuum.

\subsubsection{Envelope interpolation\label{sec:env}}
We are now left with several reliable local maxima, none or very few of them being outliers. To generate the continuum, RASSINE interpolates linearly and cubically on them, which produces two distinct continua. Some points may still be too close, producing wiggles in some cases for the reason explained in Sect.~\ref{sec:outliers} regarding item 3. %Hence, the algorithm also produces two other veils by first interpolating linearly on an equidistant grid before performing again the three previous interpolations. Such, normalisation are not necessarily better but they can be used to highlight the location where the normalisation may be unreliable.  
RASSINE also provides two other continua that can be a better match for noisy spectra, for which the upper envelope is always more or less correlated with the noise level of the spectrum. This tends to produce a continuum that is slightly too high, as already mentioned by \citet{Xu(2019)}. In this case, instead of taking the flux value of the selected local maxima to build the continuum, an average of the flux around each selected local maximum is used. The average is performed using a window of half-size \denoisdist{}, which by default is set to five wavelength elements. 

In summary, RASSINE produces four continua (linear and cubic, and denoised and undenoised) and the users are free to decide which of them they use. In Sect.~\ref{sec:precision} we show that in most cases, the linear interpolation gives the best results, therefore we recommend to use this continuum. For some applications, it might be interesting to have a continuum that can be differentiated everywhere, and thus the cubic interpolated continuum might be better. An example of the final result of RASSINE, performed in complete automatic mode (see Appendix \ref{sec:auto}), is shown for a  HARPS solar spectrum in Fig.~\ref{FigInterface2} and for four other spectra in Appendix \ref{app:collecitons}, Fig.~\ref{FigAllSpec}.

\subsection{Comparison with the AFS code \citep{Xu(2019)}}
\label{sec:xu}
A similar alpha-shape strategy for normalising spectra has been developed by \citet{Xu(2019)} with a code called AFS. It is written in R. {The authors also showed} that the alpha-shape strategy outperforms iterative methods. Interestingly enough, our codes, which are rather similar in approach, were not thought to be used on the same objects. Whereas the purpose of RASSINE initially was to normalise merged \emph{\textup{1D}} spectra, AFS was developed to remove the blaze function of the extracted echelle-order spectra. Both codes are then quite different in their conception.

Another important difference between the two codes is that in our case, the radius of the alpha shape varies according to a penalty law, which is not the case in AFS. As a consequence, even if their code contains only three parameters, they have necessarily to be fine-tuned depending on the spectral features present in the \emph{\textup{1D}} extracted echelle-order spectrum that is analysed. In comparison, RASSINE contains six parameters for the full spectrum, but five of them are calibrated and can be automatically adjusted to provide a first guess.

AFS is by nature more sensitive to broad absorption lines than RASSINE. As an example, Fig.~10 in \citet{Xu(2019)} shows that for the sodium doublet at $5890\AA$, their alpha-shape process selects the flux between the two lines as part of the continuum, whereas this location is in fact still a wing absorption. The two broad lines therefore appear slightly shallower. The same issue occurred in our case for the solar spectrum, but we resolved it by increasing the tension (see Sect.~\ref{sec:broad}). This trouble emerges in their case precisely because the authors work with \emph{\textup{1D}} extracted echelle-order spectra that are not corrected for blaze. As the interline region is precisely situated at the maximum of the blaze, the alpha shape is forced to select the region as part of the continuum. We note that \citet{Xu(2019)} described a more sophisticated correction, called ALSFS, which appears to resolve this issue. This is not clearly demonstrated in their manuscript, however.

\subsection{Normalisation of a spectral time series}
\label{sec:timeseries}
\begin{figure}[tp]
        \includegraphics[width=9cm]{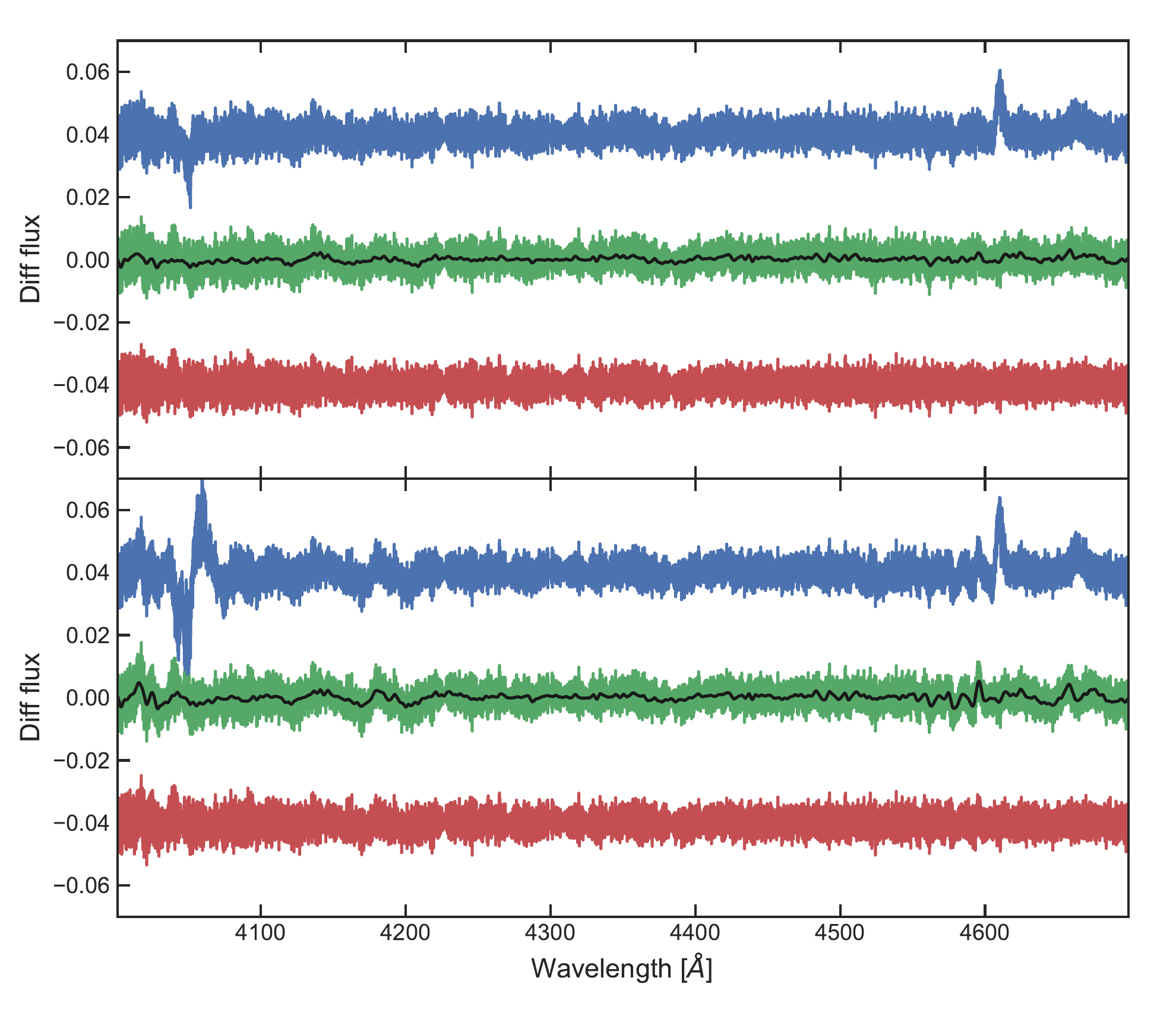}
        \caption{Visualisation of the wiggles produced in a difference of normalised spectra (blue curve) with an arbitrary zero-point. The two spectra come from the same star and are taken one night apart. \textbf{Top:} Difference between the spectra normalised by the linearly interpolated continuum. \textbf{Bottom:} Same as the top for the cubic interpolation. The green curve shows the improved stability that is reached with the clustering algorithm (see main text). The black curve shows the remaining fluctuations, and the red curve plots the residuals after those fluctuations are suppressed by the low-pass filter.}
        \label{FigWiggle}
\end{figure}

 When several spectra of the same star are considered, for instance during a series of consecutive nights, it is expected that the position in wavelength of the local maxima forming the continuum remains the same. Different local maxima from one night to the next would be due to instrumental or atmospheric effects, but not be caused by the star.  When different local maxima are used from one spectrum to the next, wiggles are induced in spectral ratios or differences, which are not desired (blue curves in Fig.~\ref{FigWiggle}). These numerical artefacts are mostly produced by the cubic interpolation (blue curve in the bottom panel of the same figure) and can already be mitigated using the linear interpolation (blue curve in the top panel of the same figure). Nevertheless, some of them are still present and need to be removed. 

To solve this problem, RASSINE provides an optional function in its library that is called \emph{intersect\_all\_continuum}. This function takes a series of spectra as input and extracts the anchor points for each of them. For each of these anchor points, the function calculates the fraction of spectra in which this anchor point has been detected. The function then selects only the anchor points that appear more frequently throughout the series of spectra than a threshold fixed with the GUI (see the screenshot of the GUI in Fig.~\ref{FigInt7}). When two maxima fall in the same cluster for one spectrum, the maximum farthest from the cluster centre is removed. Maxima that are missing from the cluster are added. In this way, we ensure that all continua are formed with the same anchor points. This considerably reduces the wiggles under the $1\%$ level (see the green curves in Fig.~\ref{FigWiggle}). We note that when the user has to work with time series of spectra at moderate S/N (from 75 up to 150), there is a risk for the clustering algorithm to fail to work properly if the smoothing step
is not efficient enough because the local maxima will be spuriously positioned. In this specific case, the user can follow the procedure described in Appendix \ref{sec:procedure}. 

%The wiggles are stronger for the cubic spline, as expected. These are induced when a local maximum (used to build the continuum curve) is selected in one spectrum but not in the others. We resolved this issue by developing a clustering algorithm on spectra time-series which can be called with the function \emph{intersect\_all\_continuum}. The resulting effect improves considerably the stability (green curve). The remaining fluctuation (black curve) are under the $1\%$ level. The residual curve after suppressing these fluctuations is shown in red and can also be obtained from the RASSINE library function \emph{matching\_diff\_continuum}.

A second function can be used, called \emph{matching\_diff\_continuum}, to reduce the wiggles even further. This function searches for the spectrum with the highest S/N at 5500 \AA\,and uses it as reference. The spectrum difference is computed with all the other spectra, one by one, the wiggles are fitted on each spectrum difference using a Savitzky-Golay filtering (black curves in Fig.~\ref{FigWiggle}) and are removed from the spectra (red curve in Fig.~\ref{FigWiggle}). The window of the filter is chosen in the GUI (see Fig.~\ref{FigInt8}). Because this step can also suppress true variations, we advise to use it with caution.  Nevertheless, this option seems to produce the most precise spectral time series (see Sect.~\ref{sec:precision} and Fig.~\ref{FigPrecision}). 

\section{Results} \label{sec:perf}

\subsection{Broad absorption lines}
\label{sec:broad}

\begin{figure*}[h]
        \includegraphics[width=18cm]{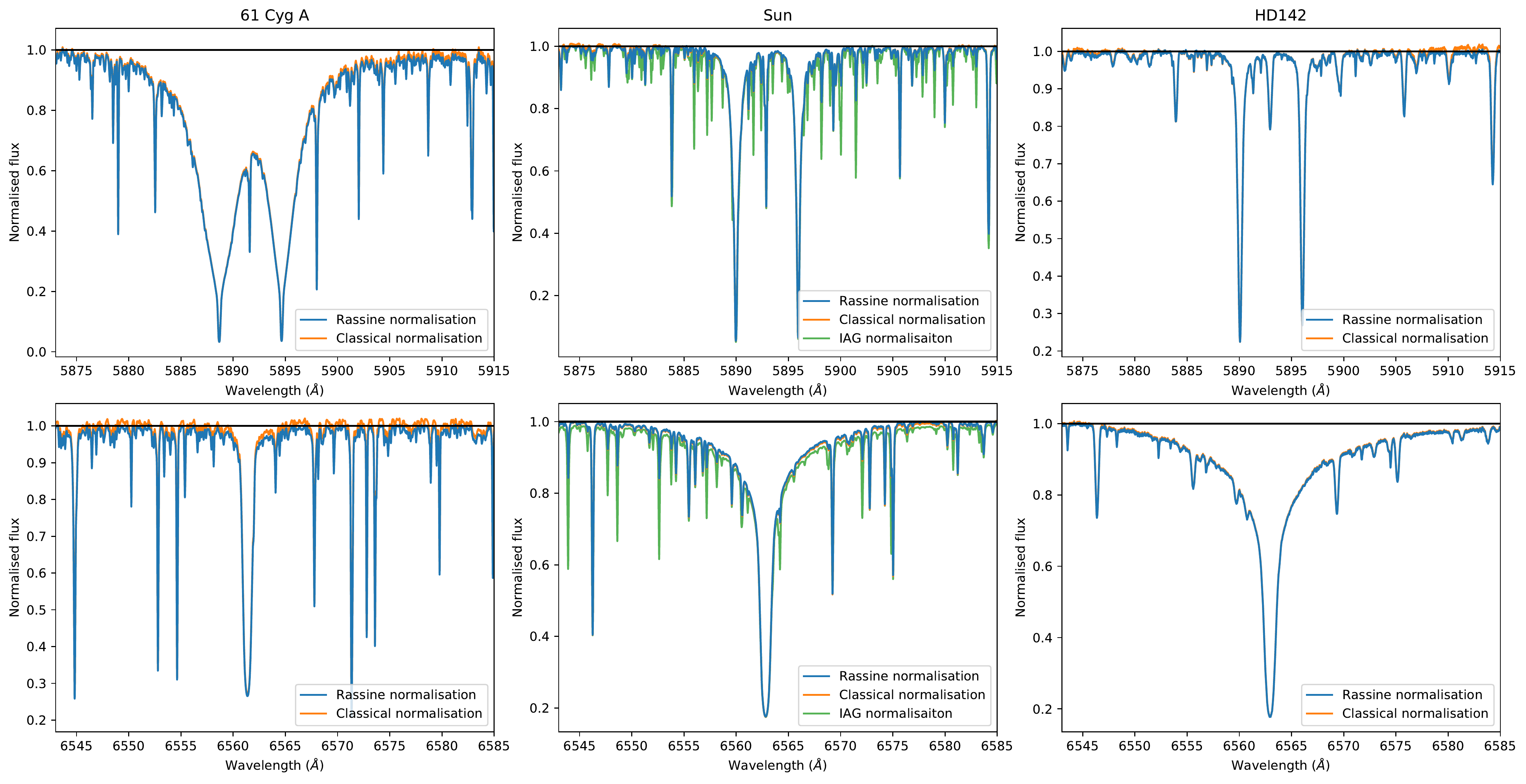}
        \caption{Comparison between the RASSINE normalisation and the normalisation obtained with a classical approach (linear fit between pseudo-continuum windows on either side of the selected line, set text) for two broad absorption lines for three stars with different spectral types: K5V, G2V, and F7V, from left to right: 61 Cyg A, the Sun, and HD142, respectively. \textbf{Top:} Spectral region corresponding to the sodium doublet NaD. \textbf{Bottom:} Spectral region corresponding to $H_\alpha$. Both 61 Cyg A and HD142 were normalised in automatic mode with intermediate tension (\pstret{} $=$ 'auto\_0.5') and polynomial mapping $\nu=1$. For the Sun, a higher tension (\pstret{} $=$ 'auto\_0.0') and smaller polynomial mapping ($\nu=0.7$) were needed to exclude the interline region of the sodium doublet from being part of the continuum. A comparison with the IAG solar atlas \citep{Reiners(2016)} degraded at the HARPS resolution $R \sim 115'000$ is also displayed. The deeper lines in the IAG atlas are tellurics.}
        \label{FigBroad}
\end{figure*} 

As a first test, we investigated whether broad absorption lines were correctly normalised when RASSINE modified their profile. To do so, we chose as test cases 61 Cyg A, the Sun, and HD142, which are three stars that probe a wide range of spectral types and CCF FWHM values (see Table \ref{TableClass} in the appendix). The spectra of 61 Cyg A were measured with the HARPS-N spectrograph, whereas the specra of the Sun and HD142  were obtained with HARPS\footnote{The Sun is observed using the HELIOS solar telescope \url{http://www.sc.eso.org/~pfigueir/HELIOS.html}}. The HARPS and HARPS-N spectra were normalised in automatic mode (see Appendix \ref{sec:auto}) with intermediate tension (\pstret{} $=$ 'auto\_0.5') and polynomial law ($\nu = 1$). The \emph{\textup{1D}} spectra of both instruments are already corrected for the blaze function, and orders are merged.

We focused on two broad lines of first interest, which are the sodium doublet NaD and $H_{\alpha}$. In Fig.~\ref{FigBroad} we compare the spectrum normalised by the alpha shape with a more classical method that is described as follows. Two spectral windows, as free of stellar lines as possible, were selected by eye on either side of each broad line, each window being different for each star. The average flux over each window was measured before the continuum was estimated as being the straight line connecting the two obtained points.

For 61 Cyg A, which is a K5V star, we observe that the classical method leads to a continuum level that is too low for the two broad absorption lines. This produces a normalised spectrum with several values higher than $1  $ because the regions that are selected as continuum for cool stars are necessarily contaminated by absorption lines because of the high density of spectral lines. The normalisation produced by RASSINE appears to be better for this star because the continuum level is higher. We note that it might still be too low regarding the true stellar continuum. %RASSINE appears thus as better normalised for this star with a continuum level higher. 

For the Sun, we observed at first a strong discrepancy of 2\% for the sodium doublet between the RASSINE continuum and the continuum derived using the classical method. This latter is compatible with the Kitt peak \citep{Wallace(2011)} and IAG \citep{Reiners(2016)} solar atlases, therefore we concluded that the normalisation was not performed correctly by RASSINE. This discrepancy was explained by the fact that the interline region between the sodium doublet was considered by RASSINE as being part of the continuum in the automatic mode. To solve this problem, we increased the tension (\pstret{} $=$ 'auto\_0.0') and decreased the coefficient $\nu$  to $0.7$ for the polynomial mapping. After doing so, the continuum found by RASSINE was similar to the continuum found using the classical method. However, our reduction leads to a smaller line width for $H_\alpha$ than is listed in IAG. This might be explained by instrument systematics, a different activity level, or by the selection of a different continuum region in which the normalisation is performed. In any case, the continuum obtained by RASSINE is similar to the continuum obtained with the classical approach on the same spectrum. 

For HD142, an F7 dwarf, a good agreement between RASSINE and the classical method is found for $H_\alpha$. However, a discrepancy is found in the continuum obtained with the classical method in the sodium doublet. A clear quadratic drift is visible in the right part of the spectra window, which is not induced by the method itself because a linear fit was performed. Such low-frequency variations are thus inherent to the merged \emph{\textup{1D}} spectrum and are either due to the instrument systematics or are produced during the construction of the merged \emph{\textup{1D}} spectrum. This latter option is more likely because this wavelength range is precisely situated in the overlapping region between the HARPS orders $56$ and $57$ (echelle orders 104 and 103, respectively).

\begin{figure*}[ht]
\centering
        \includegraphics[width=18.5cm]{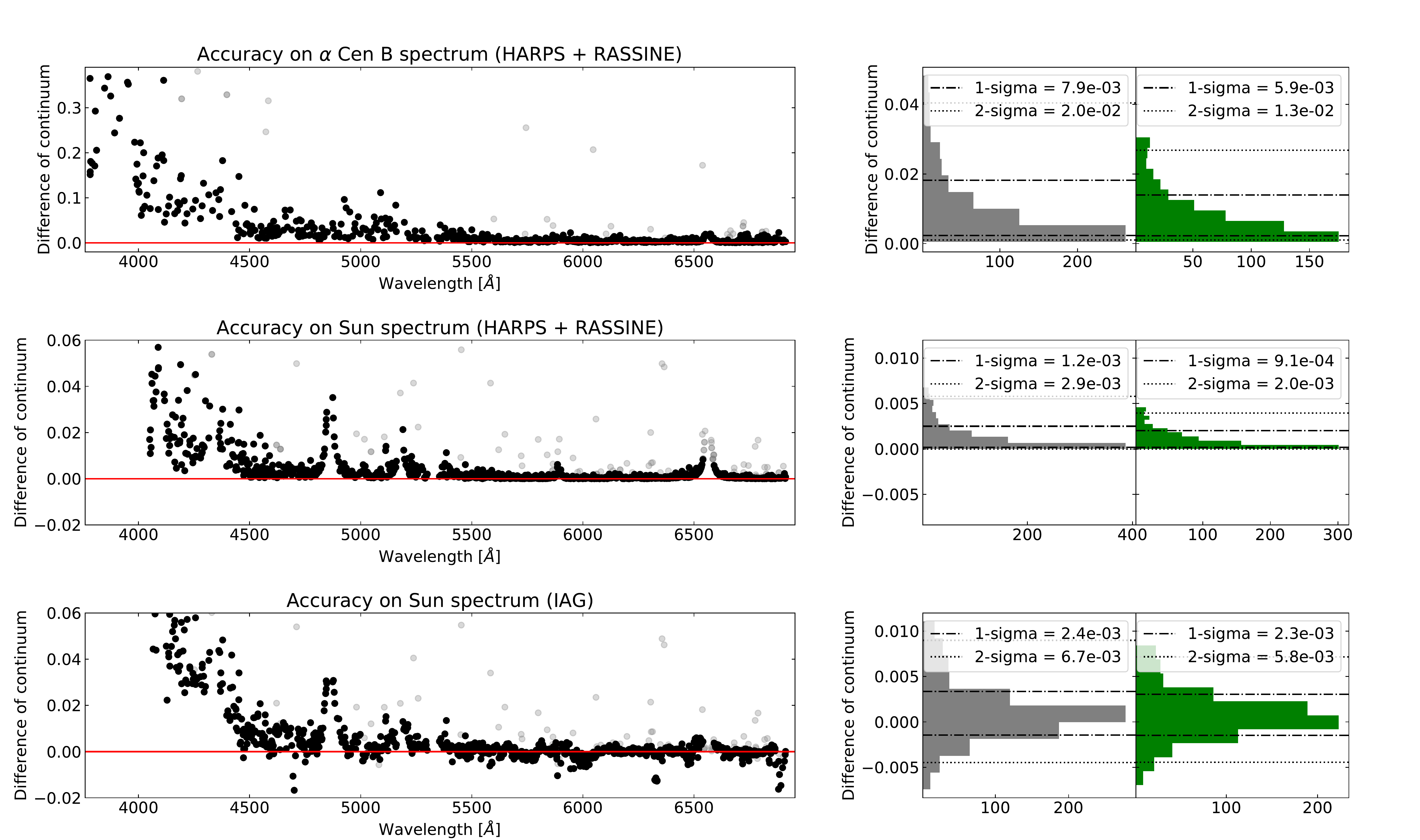}
        \caption{Computation of the RASSINE continuum accuracy by measuring the continuum difference level between our spectra and synthetic stellar templates. \textbf{First row:} Difference of the continuum for $\alpha$ Cen B in the MARCS template and the RASSINE normalised spectrum as a function of wavelength. A few continuum differences cannot be extracted correctly in the synthetic template because some stellar lines are absent from our spectra (grey dots). The distribution of the continuum difference is plotted on the right (grey histogram), as is the distribution for $\lambda>4500$ \AA\,(green histogram). 
        \textbf{Second row:} Difference of the continuum for the Sun in the ATLAS template and the RASSINE normalised spectrum.
        \textbf{Third row:} Difference of the continuum for the Sun in the ATLAS template and the IAG normalised solar atlas. The IAG solar atlas was normalised with a low-order polynomial fit.}
        \label{FigAcc}
\end{figure*}

\subsection{Evaluation of the continuum accuracy of RASSINE}
\label{sec:acc}
Even though it is not the primary interest of the RASSINE code, we tested whether the normalisation returned accurate line depths. Because the alpha-shape method implemented in RASSINE is an upper-envelope approach, it is known that for low S/N spectra the continuum level may be too high \citep{Xu(2019)}, which will provide overestimated line depths. Obtaining accurate line depths is a difficult task. Probably the best method is to rely on stellar templates. However, these templates have been generated using a few atmospheric parameters that are themselves derived from spectra that have been normalised in some way. Another strong difficulty is obtaining a stellar template with atmospheric parameters as close as possible to the observed star. Because of the high dimensionality of the stellar atmospheric parameter space, template libraries often mainly focus on a good coverage of the effective temperature and gravity surface parameters, at the expense of abundance diversity. Moreover, the resolution of an observed spectrum is generally not uniform, which might introduce incorrect line depths because a stellar template is convoluted with a kernel at a fixed resolution. As an example, the resolution for HARPS typically varies between $95'000$ and $110'000$. For this reason, the continuum accuracy cannot be computed by measuring a difference in line depth.% as we will do for the line-depth precision derivation in the next Sect. (see Sect.~\ref{sec:accuracy}).\\

To measure the continuum accuracy, we compared a spectrum normalised by RASSINE,  using the linearly interpolated continuum, with a normalised synthetic spectrum. We used the POLLUX database \citep{Palacios(2010)}, which is a library of spectra containing normalised high-resolution stellar templates. We evaluated the RASSINE continuum accuracy on two stars: $\alpha$ Cen B, and the Sun. We used the spectra from the $\alpha$ Cen B 2008 data set, but the same results are obtained with the 2010 dataset. The atmospheric parameters chosen for the templates are given in Table \ref{TableClass2} and were derived from MARCS \citep{Laverny(2012)} and ATLAS \citep{Kurucz(2005)} models. For the Sun, it is also possible to use the IAG solar atlas \citep{Reiners(2016)} in order to compare the continuum accuracy of the alpha-shape strategy with an iterative polynomial fitting method. The solar atlas and the synthetic spectra were degraded at the HARPS resolution, shifted to match the rest frame of the stellar spectra, and interpolated on the same wavelength grid. Because the IAG spectrum begins at $4047\,\AA$, we removed the shorter wavelength range from the HARPS spectrum. Conversely, because the HARPS spectrum ends at 6910 \AA, we suppressed the longer wavelength in the IAG atlas.

To measure the continuum accuracy, we extracted flux values $f_i$ of the normalised stellar template, or IAG atlas, at the same wavelength positions as the anchor points we used to build our continuum. As by definition our continuum takes a value of unity at these locations, the standard deviation of the continuum difference $1-f_i$ provides a good metric for the RASSINE continuum accuracy, which is defined here as the 2$\sigma$ width of the standard deviation distribution. We note that because this metric is only based on the anchor points (selected local maxima), no conclusion can be reached about the accuracy value of the continuum between anchor points, and thus the accuracy for the linear and cubic interpolated continuum are the same. Considering that the continuum is a smooth function between two anchor points, which is a good approximation here as the average distance between consecutive anchors points is small ($\sim4.5\,\AA$), the derived continuum accuracy can be considered as an average value representative of the alpha-shape strategy for the whole spectrum between $4047$ and $6910\,\AA$. In addition, this metric is insensitive to a global offset between our continuum and the continuum from the template because it relies on computing the standard deviation of the distribution, but this is strongly unlikely for high S/N spectra.
%extracted as the sigma-width of the distribution, taking from the percentile definition. The 1-sigma width being equivalent to the half distance between the $84^{th}$ and $16^{th}$ percentiles. The 2-sigma width is defined in the same way with the $97.5^{th}$ and $2.5^{th}$ percentiles and will be the default value when we will refer to \emph{accuracy}

Additional spectral lines were sometimes present in the synthetic spectrum and thus produced easily identifiable outliers in the continuum difference (see the grey dots in Fig.~\ref{FigAcc}). We rejected these points by performing a rolling median on 20 adjacent points, and rejected all points in the residuals that were ten times the median absolute deviation (MAD) farther away from zero. This process was performed iteratively until no more outliers were detected. Furthermore, because the IAG atlas was taken at a different barycentric Earth RV (BERV) value, a few local maxima were situated in telluric lines. We removed these points by performing the same outlier rejection. 

\renewcommand{\arraystretch}{1.6}
\begin{table}[ht]
        \caption{Atmospheric parameters chosen to generate our reference synthetic spectra. Spectra were generated either by a MARCS or ATLAS model specifying an effective temperature $T_{eff}$ in $K$, a surface gravity $\log g$, a metalicity [Fe/H], and [$\alpha$/Fe] value and a micro-turbulence velocity $\xi_{t}$ in \kms{}.}            
        \label{TableClass2}     
        \centering                         
        \begin{tabular}{ccccccc}
                \hline\hline
                
                Star & Model & $T_{eff}$ & $\log g$ & [Fe/H] & [$\alpha$/Fe] & $\xi_{t}$  \\
                % table heading 
                \hline    
                $\alpha$ Cen B & MARCS & 5250 & 4.5 & 0.25 & 0 & 1 \\
                Sun            & ATLAS & 5800 & 4.5 & 0    & 0 & 2 \\   
                \hline
        \end{tabular}
\end{table}

Figure~\ref{FigAcc} shows that our distribution of the continuum difference is always positive, meaning that the continuum fitted by RASSINE remains always below the synthetic continuum, in absolute flux. This observation is coherent with the alpha-shape strategy because the only possibility for our continuum level to be higher than the synthetic would be that one of the anchor points used to obtain the continuum were a spurious local maximum, for example due to a cosmic hit. However, as seen in Sect.~\ref{sec:outliers}, such outliers should be removed during the cleaning process. Overall, RASSINE tends to select too many anchor points as being part of the continuum, an effect that is enhanced when the tension parameter is reduced ($par\_stretching$ increased). As an example, in the case of the continuum fitted to $\alpha$ Cen B and the solar spectrum, RASSINE considers the maximum of the inter-region between the CaII H \& K lines as being part of the continuum, which is clearly not the case in the synthetic template because of the strong absorption in this region. This discrepancy produces the strong excursion of $0.35$ around $3900\,\AA$, seen for $\alpha$ Cen B in Fig.~ \ref{FigAcc}. For the Sun, the discrepancy is as high as 0.2 at the same wavelength, but it is not shown in the same figure as the HARPS spectrum was truncated at 4047 \AA\,to match the IAG solar atlas.

A summary of the continuum accuracy statistics can be found in Table. \ref{TableClass3}. The continuum accuracy is lower for $\alpha$ Cen B in the blue part of the spectrum than for the Sun. This is due to the CH molecular band at 4300 \AA, also called G band, and to the CN violet molecular band at 3883 \AA.\,These bands are deeper in cool stars and sensitive to stellar activity \citep{Berdyugina(2003)}. The continuum accuracy for $\alpha$ Cen B is estimated to be $2.0\%$, which is six times lower than the continuum accuracy measured on the Sun, $0.29\%$. This conclusion is expected  because cooler stars present more blended lines, in addition to molecular bands, which implies that fewer local maxima probe the continuum. For this reason, we also computed the continuum accuracy without wavelengths shorter than 4500\,\AA. In this more restrictive spectral range, the continuum accuracy is $1.3\%$ for $\alpha$ Cen B and $0.26\%$ for the Sun.
%Such conclusions is of course well expected since cooler the star is, more blended the lines are and thus less local maxima really probe the continuum. \comjer{comm pour moi-même : A CHECKER} \\

When the continuum of the IAG solar atlas, which is fitted using a low-order polynomial fit, is compared with the synthetic continuum, the distribution of the differences gives values above and below zero, meaning that sometimes the flux continuum level is higher and sometimes lower than the reference. The continuum accuracy we found is twice lower that the accuracy obtained with RASSINE, which is 0.67\% compared to 0.29\%. When wavelengths shorter than 4500 \AA\, are excluded, the continuum accuracy is 0.58\%, which is three times lower than the 0.20\% obtained with RASSINE. Finally, a clear excursion of 3\% around $H_\beta$ at 4861 \AA\, is visible in both IAG and HARPS spectra, which indicates that the difference is induced by the synthetic template. It does not model this spectral feature properly.

\begin{figure}[tp]
        \centering
        \includegraphics[width=9cm]{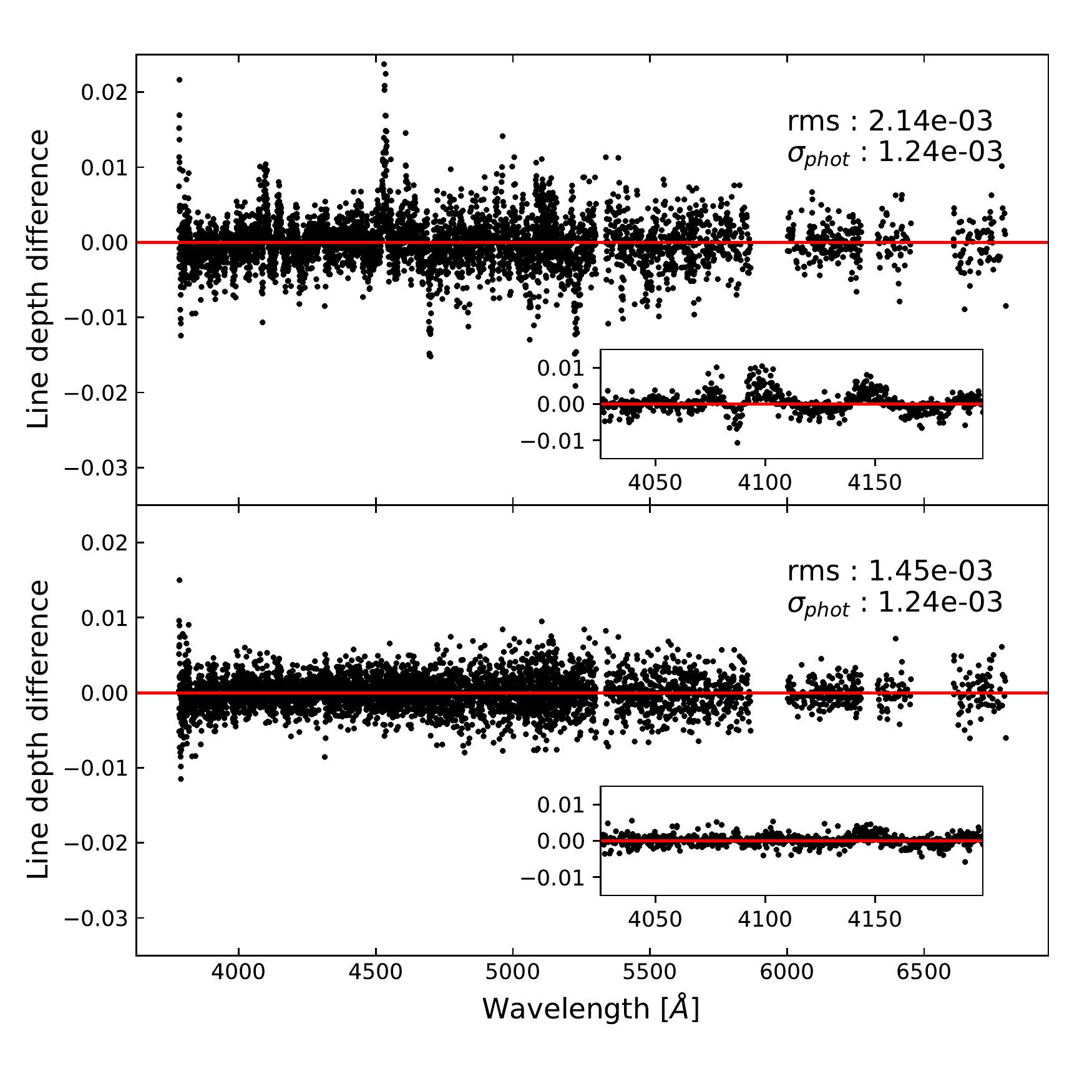}
        \caption{Line-depth difference between two $\alpha$ Cen B spectra taken one night apart (BJD = 2454557 and BJD = 2454558) . The uncertainties from the photon noise are not displayed for clarity, but their median value $\sigma_{phot}$ is indicated, as is the weighted RMS used to measure the dispersion. Both spectra are normalised by the RASSINE cubic continuum, which is the least precise continuum (see Fig.~\ref{FigPrecision}). \textbf{Top:} Line-depth difference when both spectra are normalised by the cubic continuum. This result matches the $\text{eighth}^{\text{}}$  square blue data point in the left panel of Fig.~\ref{FigPrecision} and was the least precise result, with a line-depth precision of 0.21\%. Part of the jitter is not Gaussian, as displayed in the inner plot, which is a zoom at $4100\,\AA,$ where the discrepancy is observed to be about $1.0\%$. The structures observed are induced by the cubic spline interpolation. \textbf{Bottom:} Same as the top panel after performing clustering and low-frequency filtering (see Sect.~\ref{sec:timeseries}). This result matches the $\text{eighth}^{}$ square red data point in the right panel of Fig.~\ref{FigPrecision}. The remaining jitter of 0.15\% is dominated by the interference pattern (see Fig.~\ref{FigPattern}).}
        \label{FigDiffNight}
\end{figure} 

\subsection{Evaluation of the line-depth precision of RASSINE}
\label{sec:accuracy}
Another important statistics to estimate is the precision on line depth that the code provides. For this purpose, we normalised several spectra with very high S/N of $\alpha$ Cen B that were taken during successive nights. We selected consecutive 9-night binned spectra at minimum activity in 2008, from BJD = 2454550 to BJD = 2454558. We also selected 13 consecutive nights in 2010, from BJD = 2455288 to BJD = 2455301, keeping in mind that this data set is known to be contaminated by a large magnetically active region \citep{Dumusque(2015)}, which might significantly change the depth of spectral lines \citep{Thompson(2017), Wise(2018), Dumusque(2018)}. Because we study night-to-night variations in line depth and because the rotational period is about 36 days \citep{Dewarf(2010)}, we expect the change induced by stellar activity to be negligible. This is confirmed below because both data sets give the same precision on line depth. We also selected 11 consecutive days of HARPS solar spectra from BJD = 2458507 to BJD = 2458518.

\begin{figure*}[tp]
        \includegraphics[width=18cm]{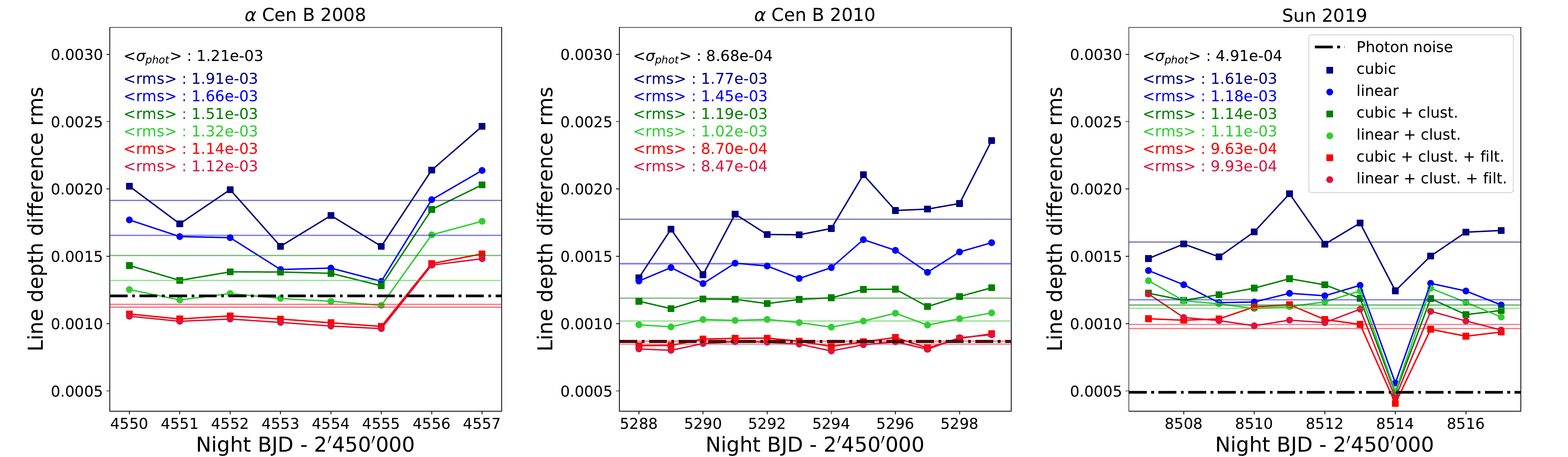}
        \caption{Measurement of the RASSINE line-depth precision using series of adjacent nights. The date of the first night forming a pair is indicated on the $x$ -axis. For each night, the line depth was computed on the full spectrum, rejecting telluric regions. The weighted RMS (coloured dots) was extracted from the distribution of the line-depth difference, where the difference is taken between two adjacent nights (see e.g. Fig.~\ref{FigDiffNight}). The average RMS $\langle rms \rangle$ on the night-to-night differences is measured and shown as horizontal solid lines. 
        %The values of $\langle rms \rangle$ for each reduction is reported in the caption. 
         \textbf{Left:} 2008 $\alpha$ Cen B dataset. \textbf{Middle:} 2010 $\alpha$ Cen B dataset. \textbf{Right:} Solar dataset. }
        \label{FigPrecision}
\end{figure*}

All the nightly stacked spectra have an S/N at $5500\,\AA$ that is higher than 1000.
%ranging from 1050 to 2390, and from 1951 to 5013. 
For each of these spectra, line depths were measured and compared between adjacent days. Because spectra are close in time and the HARPS spectrograph is extremely stable on this timescale, the line depths are not expected to change from one night to the next. The measured variation therefore is a direct measurement of the line-depth precision of RASSINE. The spectra were reduced in complete automatic mode with a Savitzky-Golay filtering, a \psbox{} of six wavelength elements, a polynomial penalty mapping ($\nu=1.0$), and using an intermediate tension (\pstret{} $=$ 'auto\_0.5').

Line depths were estimated by fitting a parabola on the points present inside a $\pm 2.5\,\kms$ window around the core of each spectral line. The uncertainty in depth was derived considering photon noise. For $\alpha$ Cen B, we fitted the lines defined in \citet{Cretignier(2019)}, and all lines contaminated by a telluric by more than 2\% were rejected. The telluric spectrum we used was generated by Molecfit \citep{Smette:2015aa} as described in \citet{Cretignier(2019)}.  For the Sun, we fitted the lines defined in the HARPS G2 mask used for cross correlation.

Two examples of line-depth differences between two adjacent nights are displayed in Fig.~\ref{FigDiffNight}. The line depth precision is computed by measuring the weighted RMS of the line-depth differences and displayed in Fig.~\ref{FigPrecision}. Because of the poor constraint of the interpolation at the edges of the spectrum, the weighted RMS was computed excluding the first and last $20\, \ang$ of the HARPS spectra. The photon noise is higher in 2008 than in 2010 and induces an average systematics on the line-depth measurement of 0.12 \% and 0.09 \%, respectively, whereas for the Sun, this is 0.05 \%. A summary of the line depth precision can be found in Table. \ref{TableClass3}. The line-depth precision is lower for the cubic interpolation than for the linear interpolation. For the former, the RMS significantly changes depending on the pair of nights that is studied, which is related to the number of wiggles that is produced in the spectral difference. On average, the RASSINE precision in line depth for the linearly interpolated continuum is 0.17 \% and 0.15 \% for 2008 and 2010, which is approximately 50\% higher than photon noise, respectively. A slightly better line depth precision of 0.12 \% is obtained for the Sun. %After measuring the quadratic difference, $D_2$, with the photon noise, the RASSINE uncertainty is found to be of $1.13\cdot 10^{-1}$ \% and $1.14\cdot 10^{-1}$ \% which is a good agreement between both datasets.  

As shown in Fig.~\ref{FigWiggle} and in the top panel of Fig.~\ref{FigDiffNight}, the continuum fitted by RASSINE presents correlated noise. These structures are produced when a local maximum is selected in a zone where no local maxima were detected for the other spectrum (see Sect.~\ref{sec:timeseries}), and this constitutes a major limitation for achieving high line-depth precision. We discuss ways below to improve this line-depth precision further.

\subsection{Evaluation of the line-depth precision of RASSINE for a spectral time series \label{sec:precision}}
 We so far investigated the line-depth precision and continuum accuracy of RASSINE for the normalisation of individual spectra. Each spectrum was thus reduced independently, without sharing any information between the different nights. However, when all the spectra are obtained from the same star and thus form a spectral time series, it is possible to gather the information of all continua to improve the normalisation. We have presented this aspect in Sect.~\ref{sec:timeseries}.
 
 %\begin{figure}[h]
%\centering
%       \includegraphics[width=9cm]{Diff_n2n_dis.pdf}
%       \caption{Same as Fig.~\ref{FigDiffNight} for the last data point of the 2008 dataset derived from the linear interpolation and after performing the clustering and low frequency filtering ($8^{th}$ red circle data point top panel Fig.~\ref{FigPrecision}). No structure are visible but a slight broader dispersion is observed at the center, the most likely explanation would be a small contamination by the companion Cen A.}
%       \label{FigDiffNight2}
%\end{figure} 
 
 We performed the previous analysis of the line-depth measurement on the 9- and 13-night binned adjacent spectra of the 2008 and 2010 $\alpha$ Cen B data sets, and also on the 11-day binned adjacent solar spectra, and this time, we applied the \emph{intersect\_all\_continuum} and \emph{matching\_diff\_continuum} functions on the three data sets separately. A summary of the line-depth precision we obtained can be found in Table. \ref{TableClass3}. 
 
As described in Sect.~\ref{sec:timeseries}, the first function  selects only local maxima that are present in most of the individual spectra, whereas the second function applies a low-pass filter to the difference between all the spectra and a reference, and individually corrects all spectra for the filtered signal (see Fig.~\ref{FigWiggle}). The precision in line depth that is obtained after these additional corrections are implemented is displayed in green and red in Fig.~ \ref{FigPrecision}, respectively. Again, the linear interpolation appears to be more precise than the cubic interpolation because the wiggles are produced by the latter. When the clustering is performed on local maxima (i.e. the \emph{intersect\_all\_continuum function} is applied) the precision in line depth significantly improves down to $0.12\%$, and when in addition low-frequency filtering is performed (i.e. the \emph{matching\_diff\_continuum function} is applied), the line-depth precision is improved even more, down to 0.10\%. This level is compatible with photon noise for $\alpha$ Cen B. For the Sun, the line-depth precision is always higher than photon noise, except for the $\text{eighth}^{\text{}}$ night pair (BJD = 2458514 and BJD = 2458515), for which the line-depth precision is compatible with it. After investigation, it turns out that this different behaviour arises because for this pair, the same flat field was used to reduce both spectra. The other night pairs are thus limited by flat fielding, which on HARPS has an S/N of $\sim$1000, which limits the line-depth precision to 0.10\%. When this is taken into account, the RASSINE precision on line depth is similar to the photon noise.

\begin{figure}[tp]
        \includegraphics[width=9cm]{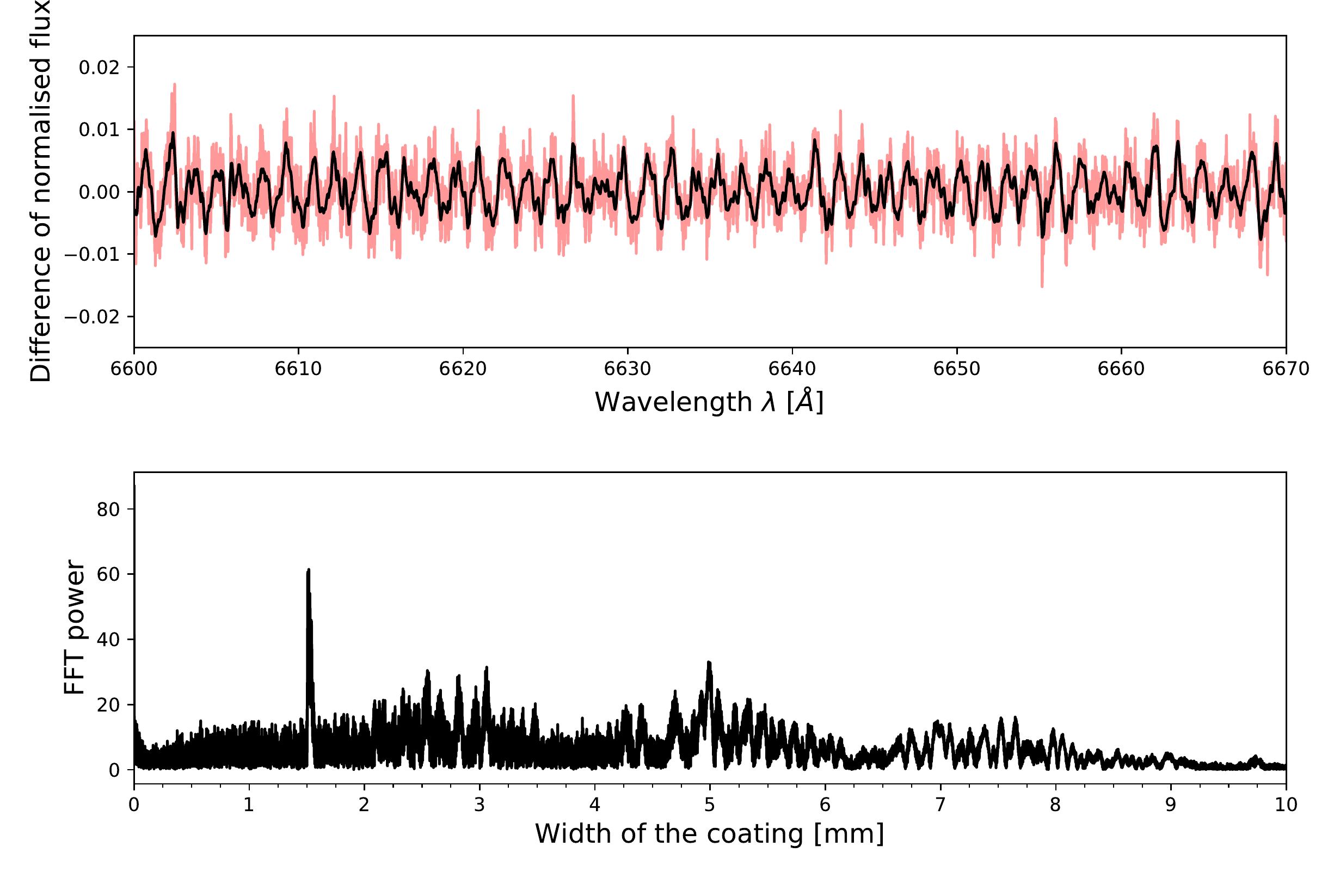}
        \caption{\textbf{Top:} Zoom into the red 6640 \AA\,region of the normalised spectral difference corresponding to the last red dot of the 2008 data set shown in the left panel of Fig.~\ref{FigPrecision}. A clear interference pattern is visible in the spectral difference (red curve) and is even more pronounced when the difference is smoothed (black curve). \textbf{Bottom:} Fourier transform of the spectral difference. A power excess is visible and corresponds to an interference pattern created by an optical element of width 1.51 mm.} 
        \label{FigPattern}
\end{figure}

In addition, we note that regardless of the algorithm used to build the continuum of the two last night pairs for $\alpha$ Cen B in 2008, the precision in line depth is clearly lower than for the other night pairs. We found that in 2008, HARPS spectra were contaminated by an interference pattern produced by a filter of 1.5mm width, that was placed in the parallel beam of the tungsten lamp that was used to perform flat fielding.  The width of the interference filter and the periodicity $\Delta \lambda$ of the observed pattern are linked by the formula $\mathrm{width} = \frac{\lambda^2}{2\Delta \lambda}$. This pattern affects the flat field and is introduced in the stellar spectrum when correcting for it. We note that this pattern remains stable for some nights and then changes significantly, producing the interference pattern in the spectral difference seen in Fig.~\ref{FigPattern}, with a peak-to-peak difference of 1\% in normalised flux units. In the data we analysed from 2008, the interference pattern remains stable for the first seven nights, and then changes for night 8 and changes again for night 9. The variation of this interference pattern during the two last nights is responsible for the degradation of the line-depth precision observed in Fig.~\ref{FigPrecision}. An upgrade of the instrument in August 2009 modified the position of this filter from parallel to diverging beam, which solved this problem. 

\section{Conclusion} \label{Conclusion}

We have presented the concepts on which the RASSINE Python code is based. The code is a tool developed to facilitate the normalisation of stellar spectra. The realisation of the code was motivated by the challenge to develop a coherent and robust normalisation algorithm of merged \emph{\textup{1D}} spectra, allowing us to deal with different spectral types and S/N, presenting a limited number of free parameters to adjust.  RASSINE proposes a GUI (see Appendix \ref{app:interface}) that allows the user to choose the best parameter values. In addition, calibrations based on the S/N level of the spectrum and the FWHM of the stellar CCF were performed to obtain first guesses for five of the six relevant RASSINE parameters (described in Appendix \ref{sec:auto}).  The code was tested with CORALIE, HARPS, HARPS-N, and ESPRESSO, four instruments that span different instrumental resolutions. The code produced visually satisfying results each time (see Appendix \ref{app:collecitons}). We focused on high S/N spectra (S/N > 500) to test the continuum accuracy and line-depth precision of RASSINE, which are thought to be the objects for which the alpha-shape strategy is expected to be efficient. This does not mean that spectra with a lower S/N cannot be reduced with RASSINE. If the smoothing step is performed correctly, the code can theoretically perform well on low S/N spectra (S/N < 100). However, the nature of the algorithm means that determining the continuum for spectra with S/N lower than 50 remains challenging.

\renewcommand{\arraystretch}{2.4}
\begin{table}[tp]
        \caption{Summary of the different statistics reached with the automatic mode of RASSINE. Only the continuum accuracy and line-depth precision obtained using the linear interpolated continuum are displayed. We also show for comparison with line-depth precision the average error induced by photon noise. The three values of line-depth precision are for individual spectra normalisation, time series of spectra normalised with clustering, and time series of spectra normalised with clustering and filtering. All the values are in percent.}            
        \label{TableClass3}     
        \centering                         
        \begin{tabular}{ccccc}
                \hline\hline
                
                Statistics & \shortstack{$\alpha$ Cen B \\ 2008} & \shortstack{$\alpha$ Cen B \\ 2010} & Sun & Sun (IAG) \\
                % table heading 
                \hline    
                Continuum accuracy & 2.0 & 2.0 & 0.29 & 0.67  \\
                \shortstack{Accuracy \\ ($\lambda > 4500 \AA$)} & 1.4 & 1.2 & 0.20 & 0.58  \\
                Line-depth precision & 0.17 & 0.14 & 0.12 & -  \\
                \shortstack{Line-depth precision \\ (clust.)} & 0.13 & 0.10 & 0.11 & -  \\
                \shortstack{Line-depth precision \\ (clust. + filt.)} & 0.11 & 0.08 & 0.10 & -  \\
                $\sigma_{phot}$ & 0.12 & 0.09 & 0.05 & -  \\
                \hline
        \end{tabular}
\end{table}

We tested the continuum accuracy and the line-depth precision of the automatic mode of the code on the 2008 and 2010 $\alpha$ Cen B data set, keeping in mind that this mode is not guaranteed to provide the best values in continuum accuracy. We showed that the linear interpolation provided a better continuum that the cubic interpolation in all cases because cubic interpolation produces undesired artefacts. The accuracy on the continuum level was measured on the 2008 $\alpha$ Cen B data set and on solar spectra by comparing our continuum with stellar templates from MARCS and ATLAS models. A continuum accuracy of 2.0\% was derived for the former star, whereas a value a 0.29\% was found for the Sun. This is 2.3 times better than the 0.67\% obtained on the IAG solar atlas, which uses low-order polynomial fitting. By considering wavelengths larger than 4500\,\AA, the accuracy of the continuum improved to 1.3\% and 0.20\%, which is three times better than for the IAG (about 0.59\%). For $\alpha$ Cen B, the line-depth precision on the 2010 data set, which does not contain the interference pattern, is about 0.14\%. Additional algorithms were developed in order to stabilise the continuum even more, in particular when the goal is to study a spectral time series. A clustering algorithm that always allows selecting the same local maxima on all spectra improved the line-depth precision down to $\sim 0.12\%$. The line-depth precision can be improved even more when a low-frequency filtering is applied on the spectral difference, which seems to produce a line-depth precision limit for RASSINE of $\sim 0.10\%$. This value was reached for $\alpha$ Cen B and solar spectra. This limit 
is the photon-noise limit on $\alpha$ Cen B spectrum and the photon-noise limit of the flat fields used to reduce solar spectra. We therefore conclude that the RASSINE normalisation was always found compatible with photon noise. All the continuum accuracy and line-depth precision statistics are summarised in Table \ref{TableClass3}.

A tool like this can find applications in numerous situations, such as computations of stellar atmospheric parameters \citep{Blanco(2019)}, the development of a tailored cross-correlation mask for each star in the context of radial velocities \citep{Bourrier(2020)}, line-by-line variability related to stellar activity \citep{Thompson(2017),Wise(2018), Dumusque(2018), Cretignier(2019)}, or correcting for the interference observed in transmission spectra. This normalisation algorithm might also improve the radial velocity derived from high-resolution spectra by reducing the jitter in radial velocity time series by providing a better colour correction and facilitating further correction algorithms for mitigating stellar activity or telluric contamination.

\section{Acknowledgments}
We are grateful to Nathan Hara and his constructive comments. We thank Lila Chergui and Yannick Demets for their help regarding English. This work has made use of the VALD database, operated at Uppsala University, the Institute of Astronomy RAS in Moscow, and the University of Vienna, and data coming from the ESO archive (Alpha Cen B) and from the HELIOS solar telescope at La Silla Observatory. This work has been carried out within the frame of the National Centre for Competence in Research “PlanetS” supported by the Swiss National Science Foundation (SNSF). M.C, J.F and R.A. acknowledges the financial support of the SNSF. F.P. greatly acknowledges the support provided by the Swiss National Science Foundation through grant Nr. 184618. X.D is grateful to the Branco-Weiss Fellowship for continuuous support. This project has received funding from the European Research Council (ERC) under the European Union\'s
Horizon 2020 research and innovation program (grant agreement No.~851555).

\bibliographystyle{aa}
\bibliography{Cretignier_2020_RASSINE}

\begin{appendix}

\section{Alpha-shape algorithm} \label{sec:as_algo}

\begin{figure}[tp]
        \centering
        \includegraphics[width=9cm]{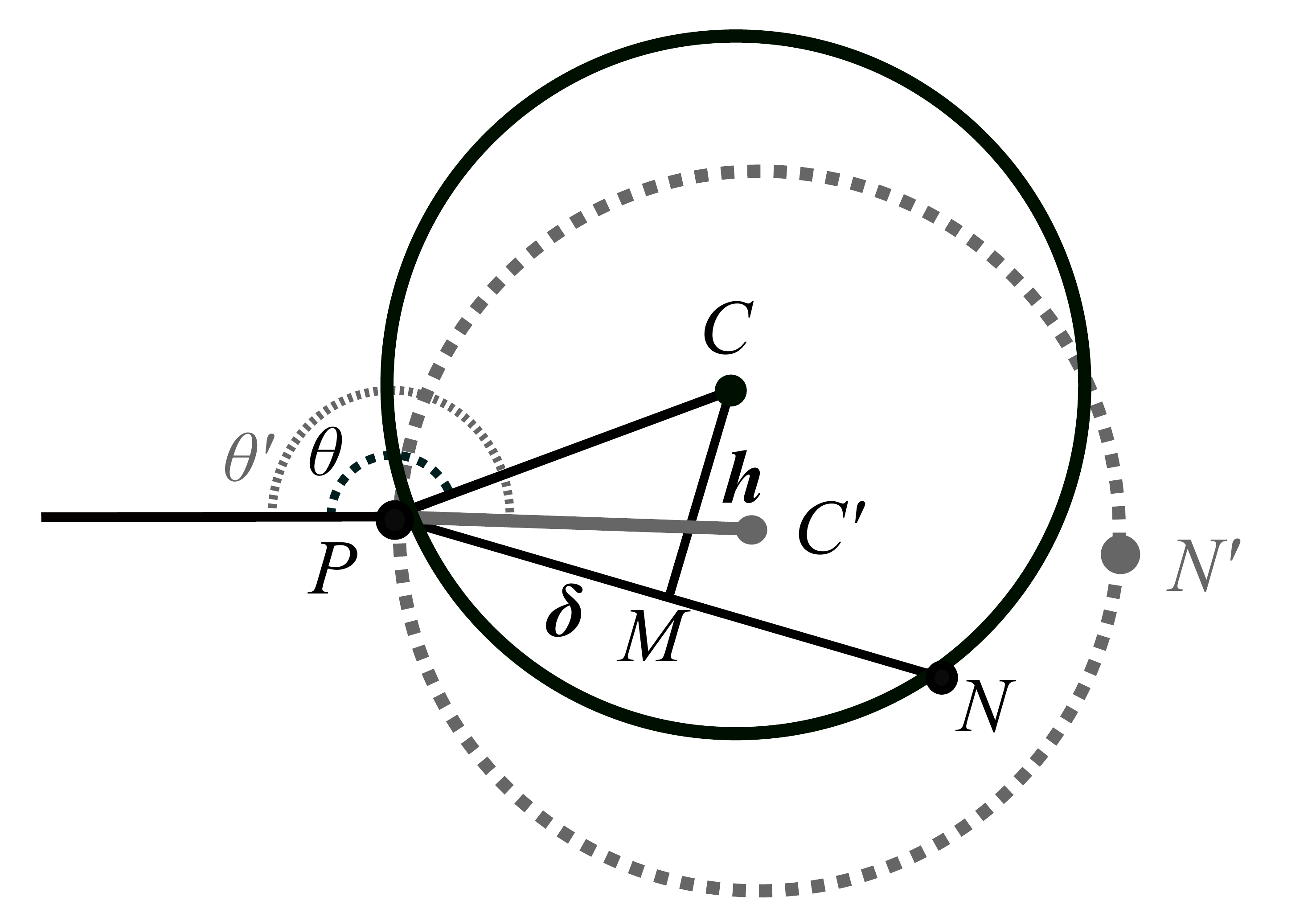}
        \caption{Schema of the gift-wrapping algorithm. The rolling pin rotates around the point $P$. $N$ and $N^\prime$ are two local maxima. The selected maximum is $N$ because the rolling pin falls on it before it falls on $N^\prime$. Mathematically speaking, this is represented by the condition $\theta < \theta^\prime$.}
        \label{FigCircle}
\end{figure}

Rolling the alpha shape, or rolling pin, on top of the spectrum given our conditions is only a problem of trigonometry. Assuming our rolling pin is situated on a local maximum (see Fig.~\ref{FigCircle}), we first obtain all the local maxima (to the right) such that the distance $d$ to the current local maximum satisfies $d<2r$. If no such points exist, the radius is increased by a factor $1.5$ until a point is found that is close enough. We call the current point $P$ and the potential next point $N$. We compute the vector from $P$ to $N$, called $\boldsymbol{\delta,}$ whose norm is $\delta$. We wish to determine the coordinates of the centre $C$ of a circle of radius $r$ (the current value) that touches $P$ and $N$. The distance $h$ between this centre and the segment $\delta$ is given by the Pythagoras theorem,
\begin{equation}
r^2 = h^2 + \frac{\delta^2}{4}.
\end{equation}
The vectorial components of the centre $C$ are then given by
\begin{equation}
\boldsymbol{C} = \boldsymbol{P} + \frac 12 \boldsymbol{\delta} + \frac{h}{\delta} \bar{\boldsymbol{\delta}}.
\end{equation}
The second term extends from $P$ to the middle of the segment $\boldsymbol{\delta}$, for instance, the point $M$. We note that $PCN$ is isosceles and $CM$ is perpendicular to $PN$. The last term of the equation is a vector perpendicular to $\boldsymbol{\delta}$ that is normalised to have norm $h$. In components terms, $\bar{\boldsymbol{\delta}}=(-\delta_y, \delta_x)$.

When the coordinates of the centre are found, we need to compute the rotation angle, that is, the angle the rolling pin has to roll to touch this particular point. Let $(p_x,p_y)$ and $(c_x,c_y)$ be the coordinates of the current point and of the circle centre, respectively. Trigonometric considerations show that this angle is given by
\begin{equation}
\theta =
\left\{\begin{matrix}
-\arccos \left( \frac{c_x-p_x}{r} \right) + \pi,& \mathrm{if} &  c_y-p_y\geq 0 \\ 
-\arcsin \left( \frac{c_y-p_y}{r} \right) + \pi,& \mathrm{if}&  c_y-p_y< 0.
\end{matrix}\right.
\end{equation}
This angle is computed for every candidate point, that is, those that are closer than $2r$ to the current point ($N$ and $N'$ in the figure). The next selected maximum is the first candidate touched by the rolling pin, mathematically speaking, the maximum with the smallest $\theta$, $N$ in this case. This process repeats iteratively until the code reaches the end of the spectrum. 

\section{Description of the automatic procedure\label{sec:auto}}

\begin{figure*}[h]
        \includegraphics[width=18.5cm]{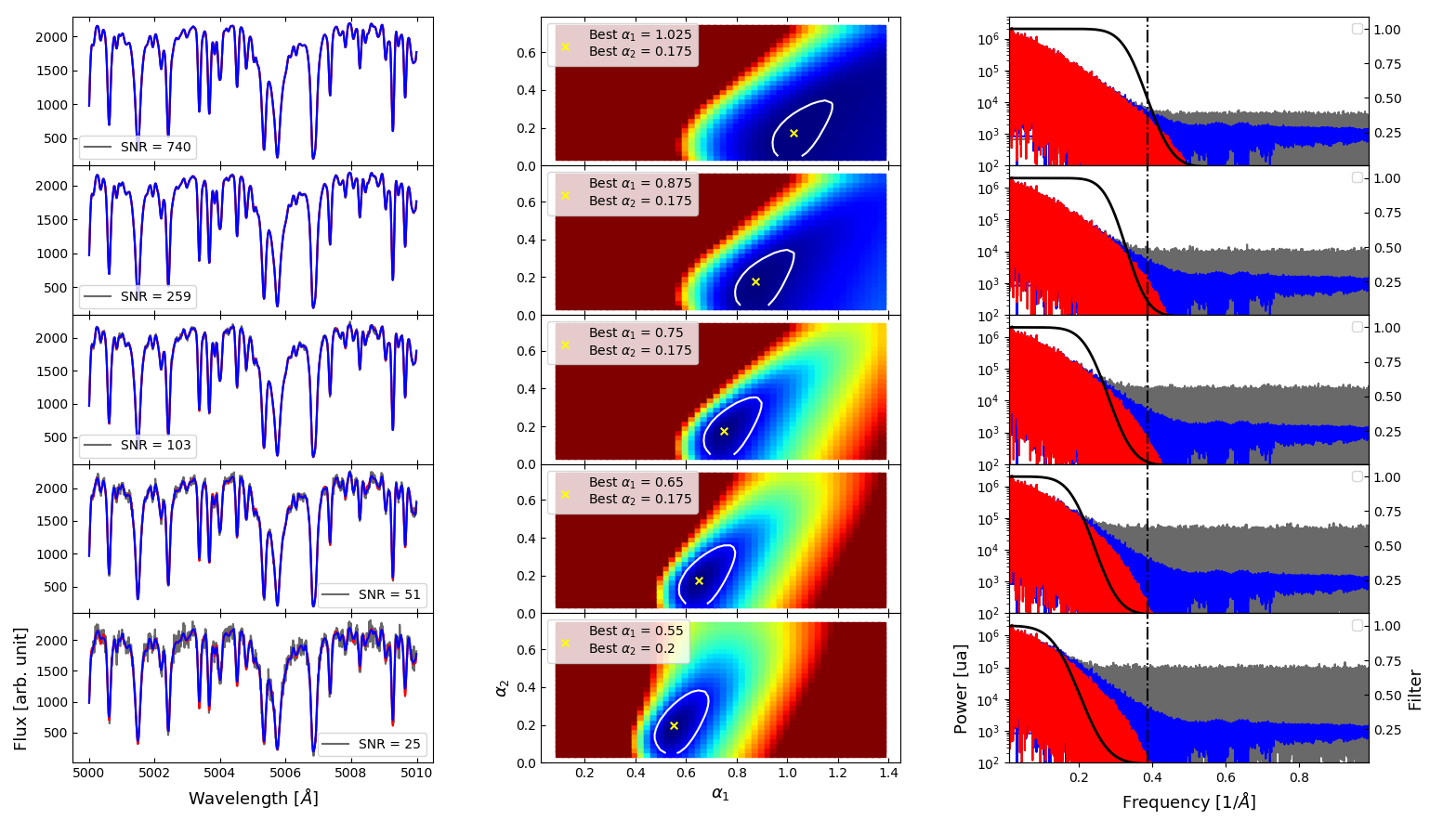}
        \caption{Calibration of $\alpha_1$ and $\alpha_2$ as a function of the spectrum S/N for the $erf$ Fourier filter. \textbf{Left:} A high S/N ($\sim$ 12000) spectrum of $\alpha$ Cen B (blue curve) is considered to be noise free. Different realisations of Poisson noise are generated (grey curves) in order to determine the best pair $(\alpha_1$,$\alpha_2)$ whose corresponding smoothed curve (red curve) is most similar to the original curve. Only a subpart of the spectra is displayed for graphical considerations. \textbf{Middle:} $(\alpha_1$, $\alpha_2)$ parameter space for different S/N realisations. The simulations were run on a grid of evenly spaced values of $\alpha_1$ and $\alpha_2$. The colours encode the metric used here, i.e. the standard deviation between the smoothed and noise-free spectrum. %The upper range colour (dark red) is limited to the median of the metric. The increase of the blue region is showing than the number of good parameters pairs increases with S/N of the spectra. 
        The local minimum of the parameter space is indicated by a yellow cross, and the contour level of the 5\% best simulations is drawn in white. The noisier the spectrum, the lower the best $\alpha_1$. The best value for $\alpha_2$ appears rather constant around $0.175$.  \textbf{Right:} Fourier transform of the spectra of the middle column (blue curve). The high-frequency noise (grey curve) increases significantly from S/N$=740$ to S/N$=25$ such that the centre of the $erf$ function moves towards shorter frequencies. The $erf$ filter corresponding to the best ($\alpha_1$,$\alpha_2$) pair is displayed (black curve). The red curve is the Fourier transform after smoothing by the function. The typical frequency scale $\sigma^{-1}$, where the sigma is the width of the Gaussian fitted to the stellar CCF, is shown as a dashed vertical line.}
        \label{FigSchemaNoise}
\end{figure*}

Even though RASSINE is a code with a complete interactive Python interface taking advantage of widgets and sliders, which facilitates choosing the parameters, it can still be technical for a user who is not familiar with this to determine the best parameters when they have several spectra of different stars of different spectral types or several spectra of the same star with different S/N. Hopefully, these parameters can often be approximated directly from the spectra. All of them, except for the penalty law, can be replaced by the keyword "auto". In this case, several algorithms that we describe below can automatically guess initial values. We recall that there is no guarantee for the quality of the final product in automatic mode. This analysis should only be used to help the user make a first guess for the continuum, or when many spectra of the same star have to be reduced. Moreover, even though theoretically nothing prevents RASSINE from being used on extracted echelle-order spectra, the user has to be aware that the automatic mode presented below was calibrated on merged \emph{\textup{1D}} spectra.

The first information that can be obtained from a spectrum is the typical line width, which is determined by the instrumental resolution, projected stellar velocity, and macroturbulence \citep{Gray(2005)}. As said previously, the easiest method to determine its value consists of computing the FWHM of the CCF, which represents roughly an average line profile. If the parameter \pccf{} of RASSINE is put in "master" mode, the code computes the CCF and its corresponding FWHM. The first step is to build the cross-correlation mask that will then be used to produce the CCF by correlating this mask with the stellar spectrum. This is done by first obtaining a crude estimate of the spectral continuum using a rolling maximum in a window of $30\, \ang$. Because the continuum we obtain has a step-like shape, we smoothed with a rectangular kernel of $15\, \ang$. We then find all the local minima with a derivative criterion on a Savitsky-Golay smoothed version of the spectrum. This local minimum corresponds to spectral lines, and the difference between these local minima and the neighbour-fitted continuum provides an approximate depth for each spectral line. The final mask is a collection of the wavelength of each minimum with its respective depth, as defined in \citet{Pepe(2002)}. The CCF obtained from cross-correlating the mask with the stellar spectrum is then fitted with a Gaussian and the FWHM is extracted. A list of wavelength bands can be given as parameters \pmask{} in order to exclude some regions that are contaminated by telluric lines from the obtained CCF mask. A mask can also be given directly as input in the \pccf{} parameter, in which case the user also has to specify the systemic velocity of the star so that the stellar spectrum and binary mask are in the same rest frame. This is more complex, and we do not recommend using this option for users unfamiliar with RASSINE. 

When the FWHM is known, the first step consists of smoothing the spectra. If the \psbox{} parameter is set to "auto", a high-frequency filter is applied in Fourier space to suppress the frequencies above the sigma width. Two filters are available: an error function, and a top hat with an exponentially decreasing tail. Their expressions are given by 
\begin{equation}
\label{eq:filters}
\begin{matrix}
f_{erf}(\omega) &=& C\ \mathrm{erf}\left(\frac{\omega_0 - \left | \omega \right |}{ \delta \omega} \right) & \\ 
%f_{FD}(\omega) &=& C  \left( 1 + \frac{1}{1+e^{ \frac{\left | \omega \right | - \omega_0}{\delta} } } \right ) & \\
\ &\ \\
f_{hat}(\omega) &=& \left\{\begin{matrix}
1 & \left | \omega \right | \leqslant \omega_0  \\ 
e^{ -\frac{\left | \omega \right | - \omega_0}{\delta} } & \left | \omega \right | > \omega_0  \\ 
\end{matrix}\right.
\end{matrix},
\end{equation}

where $C$ is a normalisation constant (the filters should all satisfy $f(0)=1$), $\omega_0$ is the centre of the filter, which can be understood as the cut-off starting at which high-frequency modes are suppressed, and $\delta \omega$ is the width of the filter (a low value of $\delta \omega$ produces a sharp filter, and a high value produces a smooth filter).

By default, the error function filter is used if the \pskernel{} parameter is set to "auto". Some care is then needed to derive the correct values for $\omega_0$ and $\delta \omega$. As mentioned before, a typical wavelength scale is given by the width of the Gaussian fitted on the CCF. This implies the existence of a given frequency scale $\sigma^{-1}$. Hence, we can parametrise $\omega_0 = \alpha_1 \sigma^{-1}$ and $\delta \omega = \alpha_2 \sigma^{-1}$, where the $\alpha_i$ values are two dimensionless parameters. To adjust them to the best values, a calibration curve was constructed using a high S/N ($\sim$ 12000) spectrum of $\alpha$ Cen B and of the Sun, which were constructed by stacking 1000 observations together. We took care of correcting the spectra for long radial velocity trends in both cases in order to allow an optimal stacking. 

\begin{figure}[tp]
        \includegraphics[width=9cm]{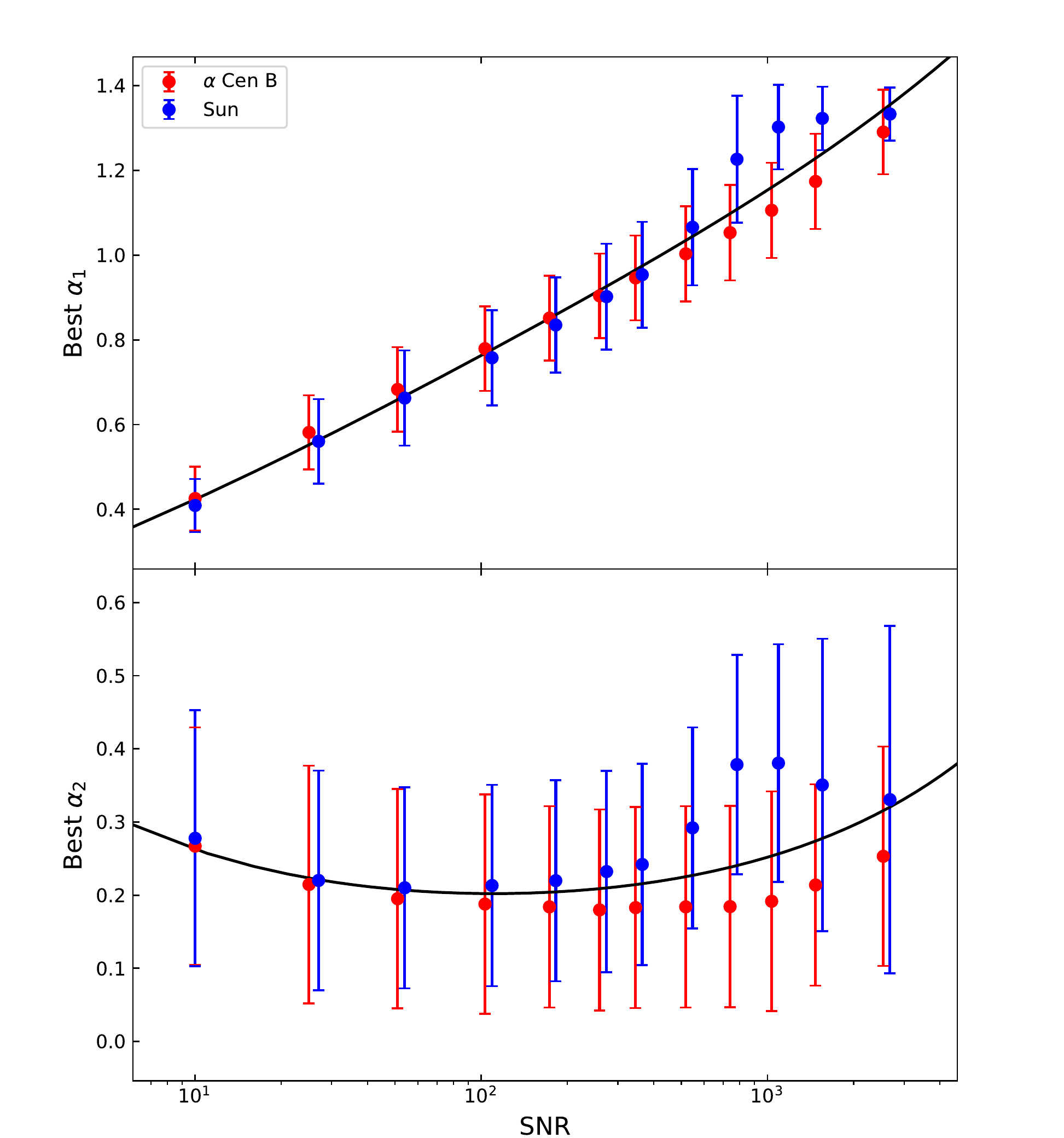}
        \caption{Calibration curves of the $\alpha_1$ and $\alpha_2$ parameters depending on the S/N value of a spectrum for the $erf$ Fourier filter. \textbf{Top:} Calibration of $\alpha_1$. The dots represent the best value according to the simulations (see Fig.~\ref{FigSchemaNoise}) that produce the least standard deviation (weighted, as explained in the text) between the noise-free and the smoothed spectrum. The error bars correspond to the $5\%$ best values of the simulations. Two stars were used for the calibration (highlighted by the colour of the points). The calibration curves were fitted by a third-degree polynomial function in log-log space. \textbf{Bottom:} Calibration of $\alpha_2$ with the same procedure. The calibration curves were fitted by a second-degree polynomial function in log-log space.}
        \label{FigSchemaS/N}
\end{figure}

We then simulated spectra with different S/N by adding several levels of Poison noise to the spectrum (see Fig.~\ref{FigSchemaNoise}). The Fourier filter was performed in an optimisation grid of $\alpha_1$ and $\alpha_2$. The best values of the $\alpha$ parameters were determined by measuring the standard deviation in the spectral difference between the noise-free and smoothed spectrum.  
By minimising this quantity, we determined the best values for $\alpha_1$ and $\alpha_2$ for each S/N value, which was extrapolated for every value of the S/N (see Fig.~\ref{FigSchemaS/N}) by fitting a polynomial function. The error bars $\Delta \alpha$ were defined as $\Delta \alpha_{1,2} = 0.5*(max(\alpha_{1,2})_j -  min(\alpha_{1,2})_j$), with $j$ the collections of the 5\%\  best simulations delimited by the white contour in Fig.~\ref{FigSchemaS/N}. The centre of the filter is more important than its width because a clear dependence is observed for $\alpha_1$ , but not for $\alpha_2$. If the automatic mode for the spectral smoothing is chosen, the flux units of the input file have to be in analogue-to-digital (ADU) units such that the S/N value can be extracted by taking the square root of the flux. We advise using this automatic smoothing mode only when several spectra that span different order of magnitude in S/N have to be reduced. 

\begin{figure}[tp]
        \includegraphics[width=9cm]{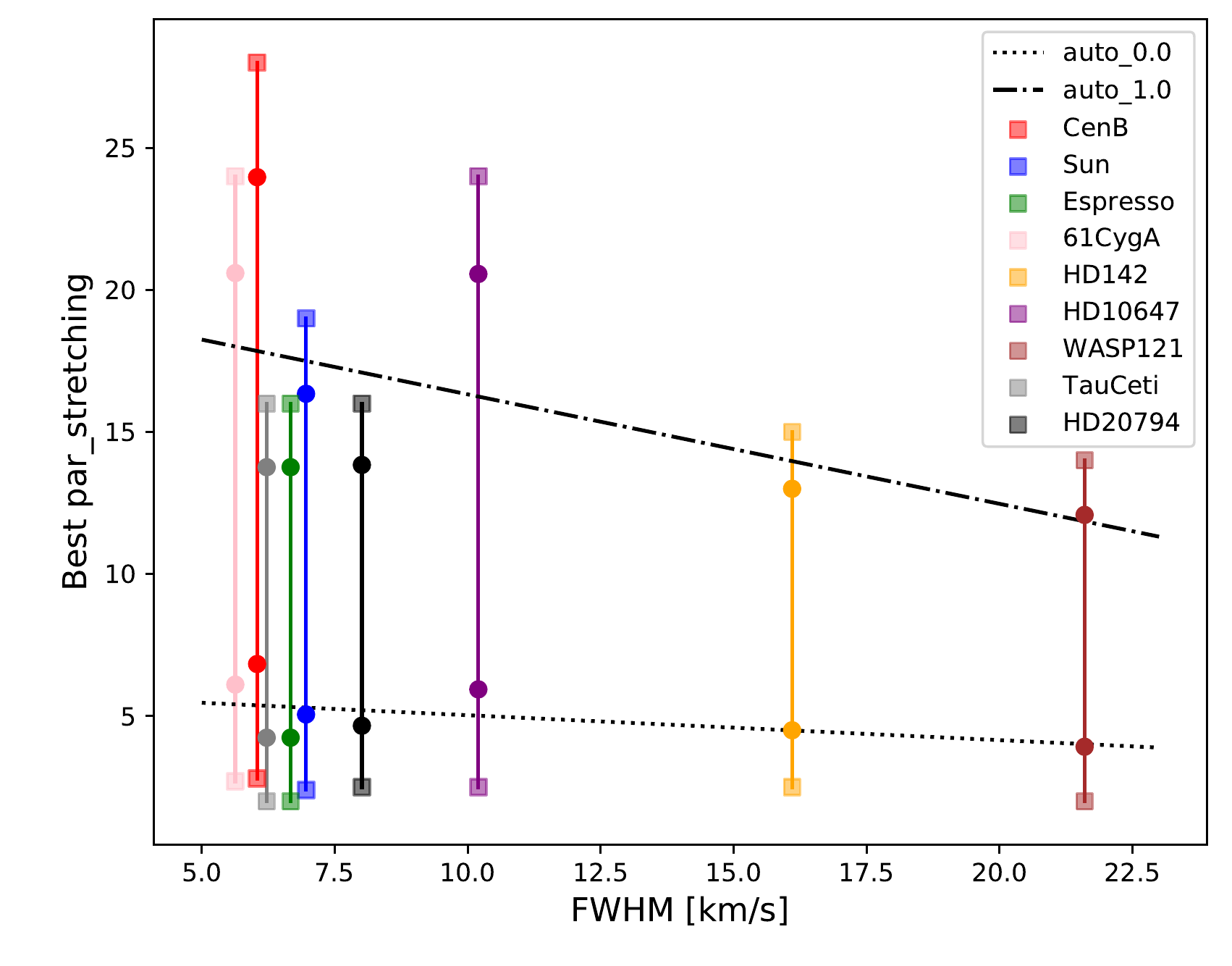}
        \caption{Calibration curves of the \pstret{} parameter depending on the CCF FWHM value derived for a given spectrum. The calibration was established $\text{based on nine}$ stars (see Table \ref{TableClass}) that probe the spectral type range from F6 to K5 (each star is indicated with a different colour). The square dots indicate the minimum and maximum values for the \pstret{} parameter (chosen by eye). The circles are more conservative ($16\%$ and $84\%$ of the parameter range). We fit the latter on a line to derive the minimum value in automatic mode (\pstret{} $=$ 'auto\_0.0', dotted) and the maximum of the parameter (\pstret{} $=$ 'auto\_1.0', dash-dotted) as a function of the FWHM. The final choice is then at the user's discretion. The nearly constant line shows the minimum value, and the decreasing line shows the maximum value.}
        \label{FigSchemaStretch}
\end{figure}

After the spectrum was smoothed, we stretched the flux units by rescaling the y-axis with respect to the x-axis, which can be done by adjusting the \pstret{}. This parameter can be compared to a tension applied on the veil that represents the stellar continuum. If the stretching parameter is too small, there is too much tension on the veil and the rolling pin can go through the spectrum, shown by the blue curve in Fig.~\ref{FigStretch} (we recall that the rolling pin considers only the local maxima, and there is no spectrum from its point of view). Similar problems can arise when the maximum radius for the rolling pin is too small. In contrast, when the parameter is too large, the tension is not strong enough and the veil starts to fall inside the lines of the spectrum, shown by the green curve in Fig.~\ref{FigStretch}.

There is a priori no precise way to determine the best value for this parameter. 
The calibration is presented in Fig.~\ref{FigSchemaStretch} and was performed by eye with nine high S/N spectra of $\text{nine}$ different stars listed in Table \ref{TableClass}. We need a typical length scale in order to calibrate the stretching parameter. The only such quantity is again the FWHM of the CCF. Furthermore, as described previously, there is a wide range of values that work quite well to normalise the spectra, depending on the amount of tension we desire. The calibration thus provides for a given value of the FWHM a range over which the parameter can be taken.
We fit two lines, one to derive the minimum value in automatic mode (\pstret{} $=$ 'auto\_0.0'), and one for the maximum value of the parameter (\pstret{} $=$ 'auto\_1.0'). The user therefore has to specify the level of tension in its continuum after the auto keyword  by also entering a number between 0 and 1. The 0 value represents a strong tension, and 1 represents a weak tension. By default, an intermediate tension is used (\pstret{} $=$ 'auto\_0.5').

\renewcommand{\arraystretch}{1.6}
\begin{table*}[tp]
        \caption{Stars and master spectra used to calibrate the \pstret{} parameters. The stars were chosen to probe different instrumental resolutions and spectral types ranging from $R \sim 55'000$ to $140'000$ and from F6V to K5V. The table contains from left to right the name of the star, its spectral type, the number of exposures stacked to form the master spectrum, the S/N of the master spectrum at 5500 \ang, the instrument that observed the star, and the average observation date. At the end we list three parameters that were determined in automatic mode by RASSINE: the FHWM of the CCF in \kms, the \pstret{} with intermediate tension (\pstret{} $=$ 'auto\_0.5'), and the \prmax{} value in \ang.}            
        \label{TableClass}     
        \centering                         
        \begin{tabular}{ccccccccc}
                \hline\hline
                
                Name & Spec. type & \# & S/N & Instr. & Date & FWHM & \pstret{} & \prmax{} \\
                % table heading 
                \hline    

                WASP-121 & F6V & 140 & 406 & HARPS & 2018/01 & 21.6 & 7.6 & 138 \\        
                HD142 & F7V & 40 & 595 & HARPS & 2004/10 & 16.1 & 9.2 & 50 \\                      
                HD10647 & F9V & 56 & 784 & HARPS & 2004/10 & 10.2 & 10.8 & 86 \\
                %Sun & G2V & 711 & 8950 & HARPS-N & 2015/08 & 6.96 & 11.7 & 78 \\                         
                Sun & G2V & 1294 & 12445 & HARPS & 2019/01 & 6.96 & 11.7 & 78 \\                         

                HD20794 & G6V & 28 & 968 & CORALIE & 2016/12 & 8.01 & 11.7 & 76 \\                 
                $\tau$ Ceti & G8V & 101 & 3036 & HARPS & 2017/07 & 6.22 & 11.9 & 76 \\    
                $\alpha$ Cen B & K1V & 1767 & 12244 & HARPS & 2010/03  & 6.07 & 11.9 & 72 \\
                $\alpha$ Hor & K2III & 2 & 1081 & ESPRESSO & 2019/01 & 6.67 & 11.8 & 72 \\  
                61 Cyg A & K5V & 129 & 3494 & HARPS-N & 2013/03 & 5.63 & 12.1 & 65 \\                    

                \hline
        \end{tabular}
\end{table*}

Figure~\ref{FigSchemaStretch} shows  interesting features. The minimum value of the parameter seems to be rather constant for all the stars, regardless of the value of the FWHM. This is the case because the flux of all the stars is normalised in the same way,  by scaling the $x-$ and $y$ -axis on the same length. We recall that when the value of the parameter is too low (high tension on the veil), the rolling pin will go through the spectrum by reaching unsuitable local maxima (corresponding to blended lines). These blended lines are present for most of the stars, either because of high stellar line density or high rotational broadening. A minimum value of 2 was therefore found to be the same lower limit for all the stars. The upper value for the parameter decreases when the FWHM increases. If the value of the parameter is too high (low tension on the veil), the code will fall into absorption lines. This means that when the value of the FWHM is already high, there is not much room to stretch the horizontal axis before we obtain lines that are broad enough to cause the rolling pin to fall. The value of the parameter therefore has to be lower as the FWHM increases.

The last two automatic parameters are the minimum and maximum radius of the rolling pin, $R$ and $R_{max}$. The former is fixed as $100$ times the $\sigma$ value of the CCF transformed in $\ang$ with the bluest wavelength of the spectrum, which prevents the rolling pin from falling inside the stellar lines. The maximum radius is computed with the same pre-continuum $S_1$ and $S_2$ as we used for the penalty (see Sect.~\ref{sec:penal}). The idea is to detect the largest absorption region, therefore we compute the difference between $S_1$ and $S_2$. $R_{max}$ is then defined as the longest cluster for which $S_1-S_2$ retains the same sign. 

\section{Spectral time series with low S/N}\label{sec:procedure}

Spectra with a low S/N are difficult data to deal with in general. For alpha-shape algorithm, we already described the risk that the upper envelope may fit the noise envelope and not the continuum, which justifies that the low-pass filter is performed as a first step on the spectra (see Sect.~\ref{sec:smooth}). Another concern might exist for the clustering algorithm presented in Sect.~\ref{sec:timeseries} in the case of spectral time series. For spectra with low S/N, the wavelength positions of the local maxima could be spuriously distributed, which contradicts the main assumption used in the clustering algorithm and renders it inefficient. 

When $N$ low-S/N spectra are to be normalised with the same anchor points, first, a master spectrum needs to be built by stacking all the individual spectra. Some care should be taken to shift them in the same rest frame.  Then RASSINE should be run on this master spectrum to normalise it and find the optimal anchor points. Finally, the function \emph{intersect\_all\_continuum} should be run, with the $N$ spectra as input, and as optional argument \emph{master\_spectrum} the name of the master spectrum. This will enforce the code to use the same anchor points as the master for each spectrum normalisation. After this, the user can run the \emph{matching\_diff\_continuum} function with the $N$ spectra and the master spectrum. This entire sequence is implemented in the code itself. 

\section{Graphical user interfaces (GUIs) of RASSINE}
\label{app:interface}

\begin{figure*}[h]
\centering
        \includegraphics[width=18.5cm]{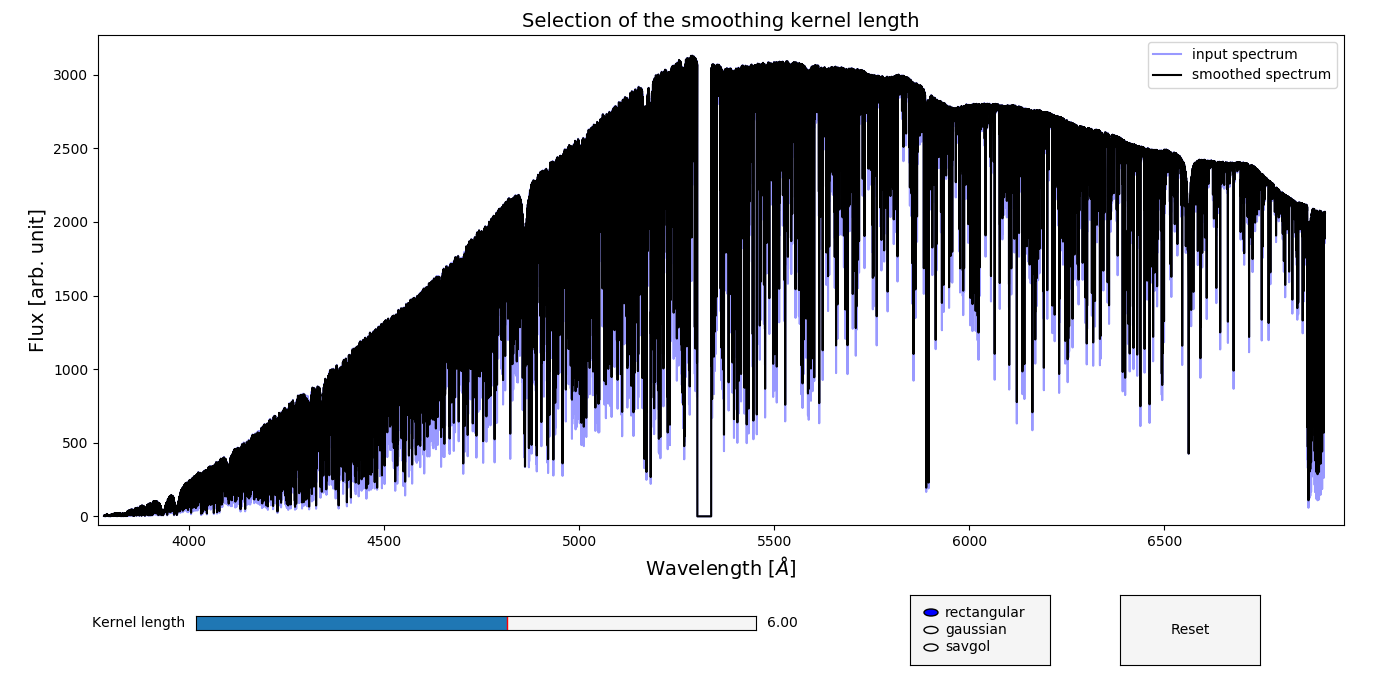}
        \caption{First GUI of RASSINE for selecting the smoothing kernel. The user has only to select the smoothing length and kernel shape.}
        \label{FigInt1}
\end{figure*}

\begin{figure*}[h]
\centering
        \includegraphics[width=18.5cm]{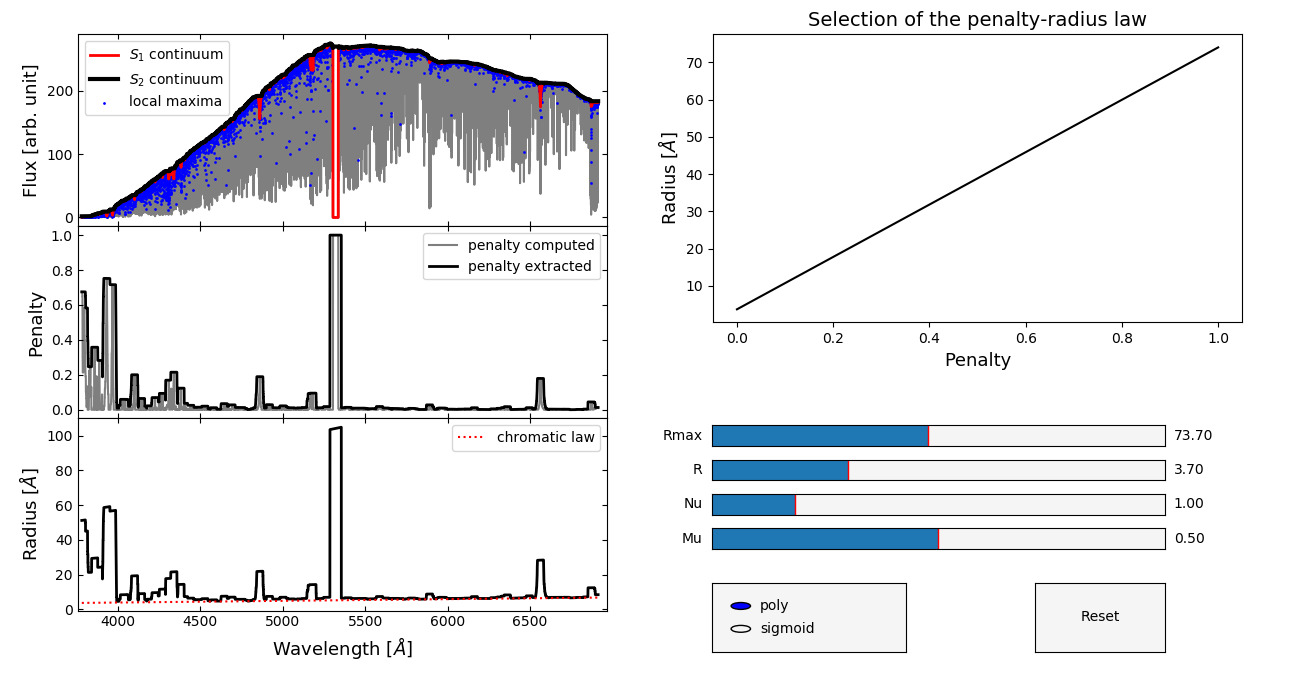}
        \caption{Second GUI of RASSINE for selecting the penalty law. The user has to select the functional form of the penalty law (polynomial or sigmoid) and the minimum and maximum values. }
        \label{FigInt2}
\end{figure*}

\begin{figure*}[h]
\centering
        \includegraphics[width=18.5cm]{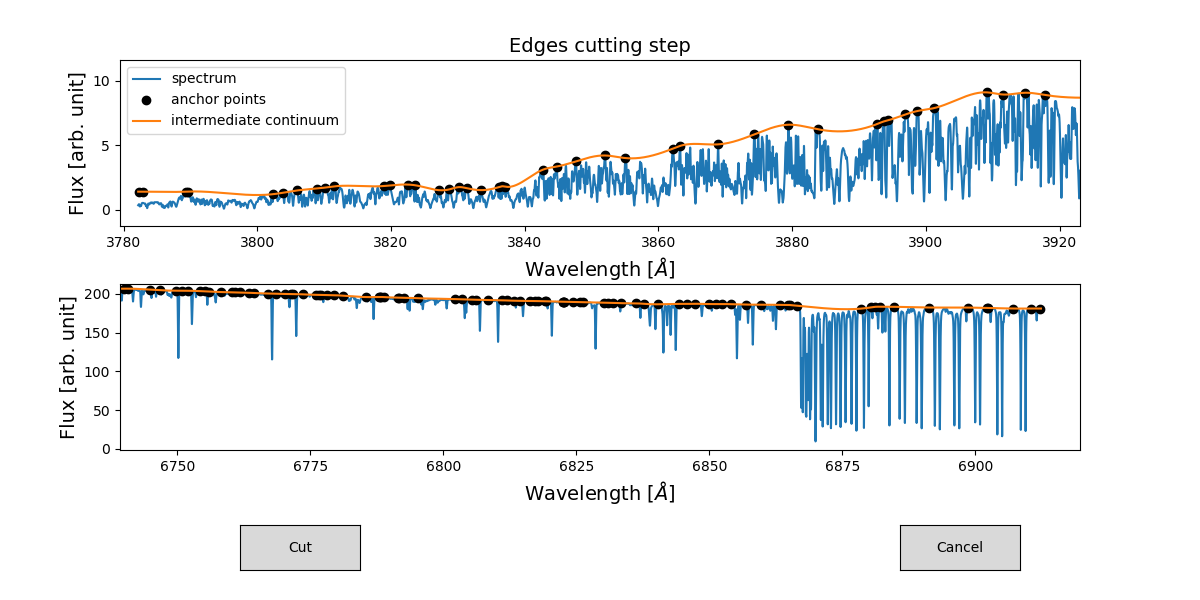}
        \caption{Third GUI of RASSINE for cutting the continuum edges. The user has to click on the "cut" button until they are satisfied.}
        \label{FigInt3}
\end{figure*}

\begin{figure*}[h]
\centering
        \includegraphics[width=18.5cm]{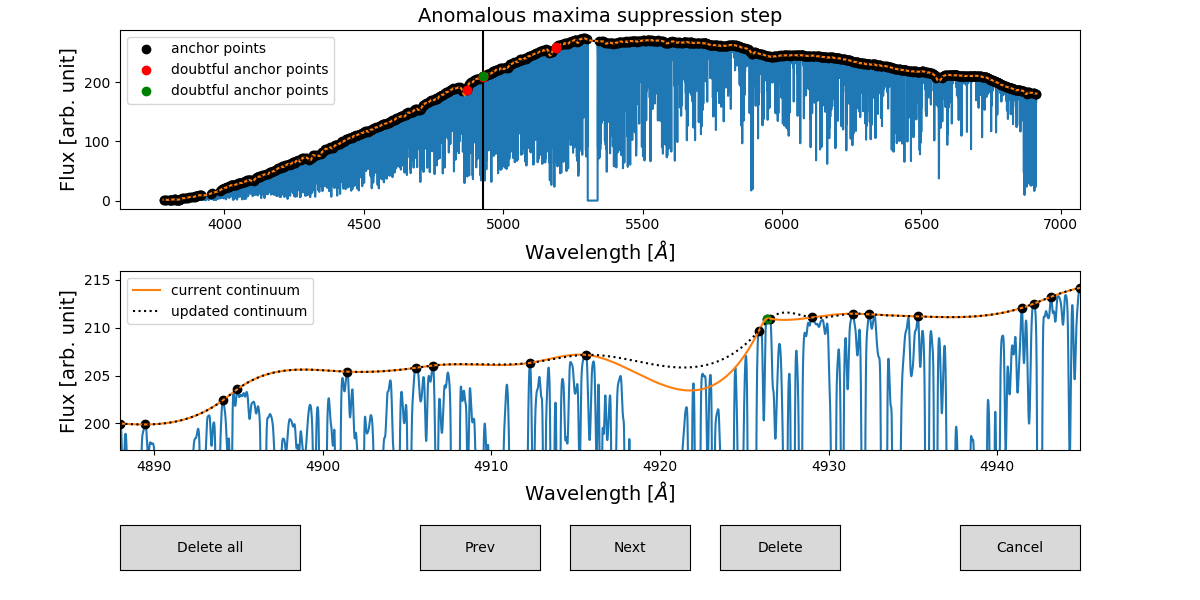}
        \caption{Fourth GUI of RASSINE for  suppressing anomalous local maxima. The user can navigate through five anomalous maxima at the same time.}
        \label{FigInt4}
\end{figure*}

\begin{figure*}[h]
\centering
        \includegraphics[width=18.5cm]{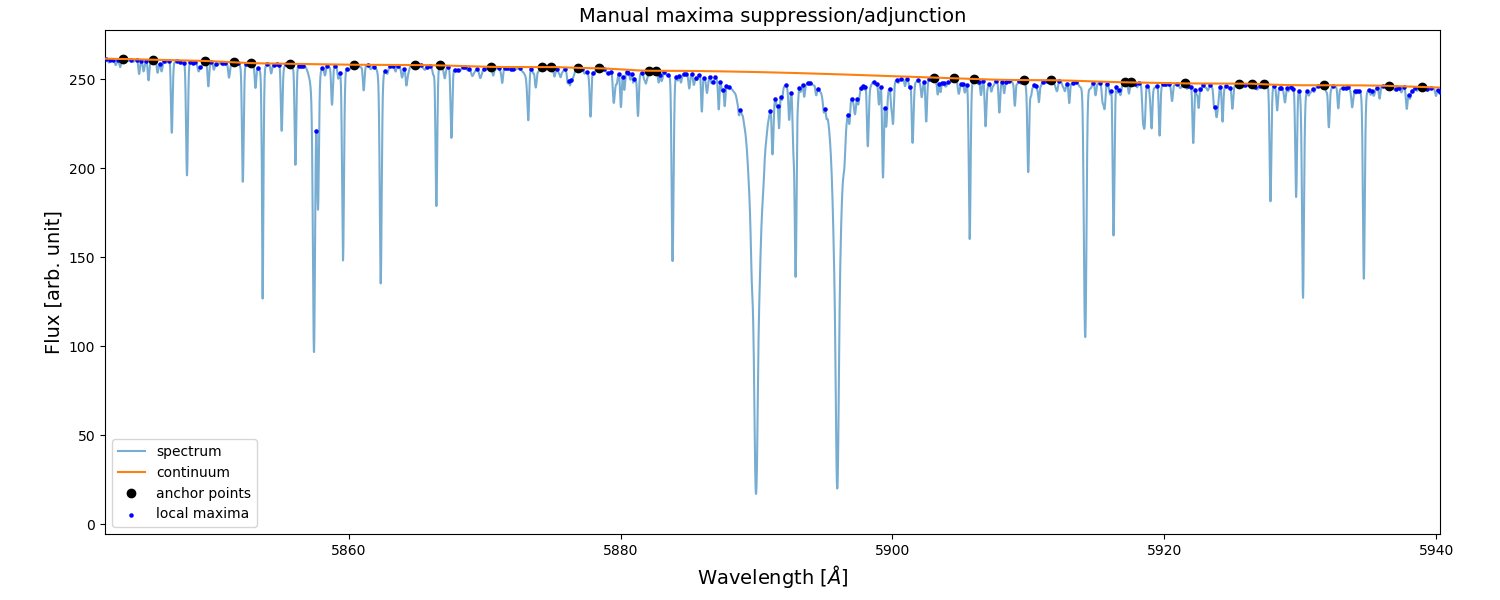}
        \caption{Fifth GUI of RASSINE for suppressing or manually adding some local maxima by clicking on the points.}
        \label{FigInt5}
\end{figure*}

\begin{figure*}[h]
\centering
        \includegraphics[width=18.5cm]{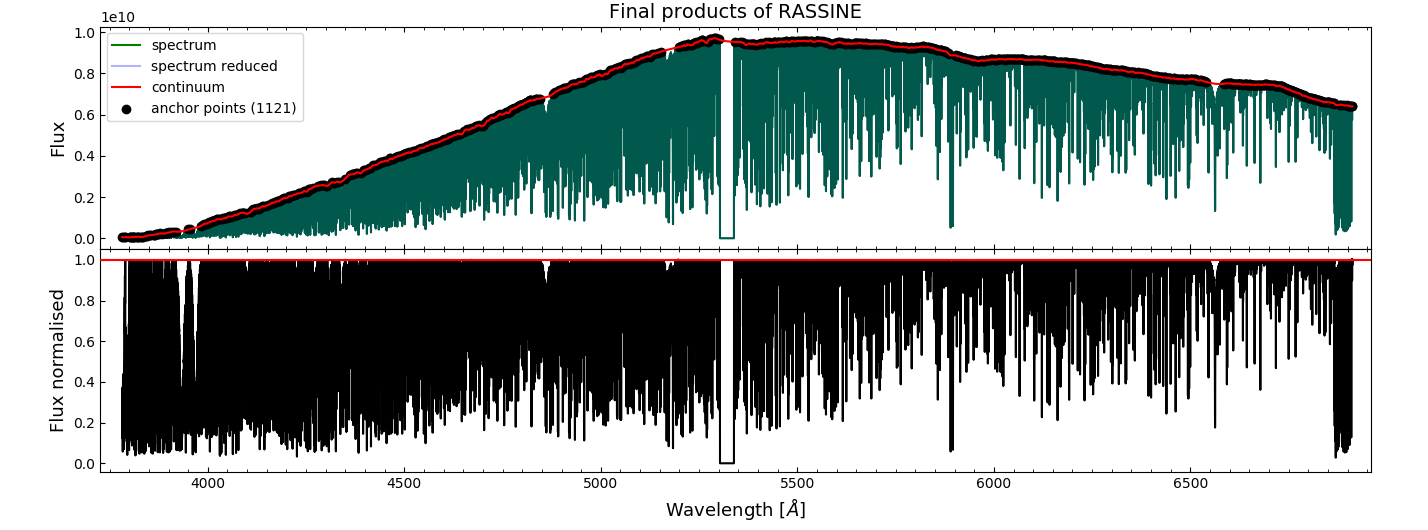}
        \caption{Last GUI of RASSINE presenting the final output.}
        \label{FigInt6}
\end{figure*}

\begin{figure*}[h]
\centering
        \includegraphics[width=18.5cm]{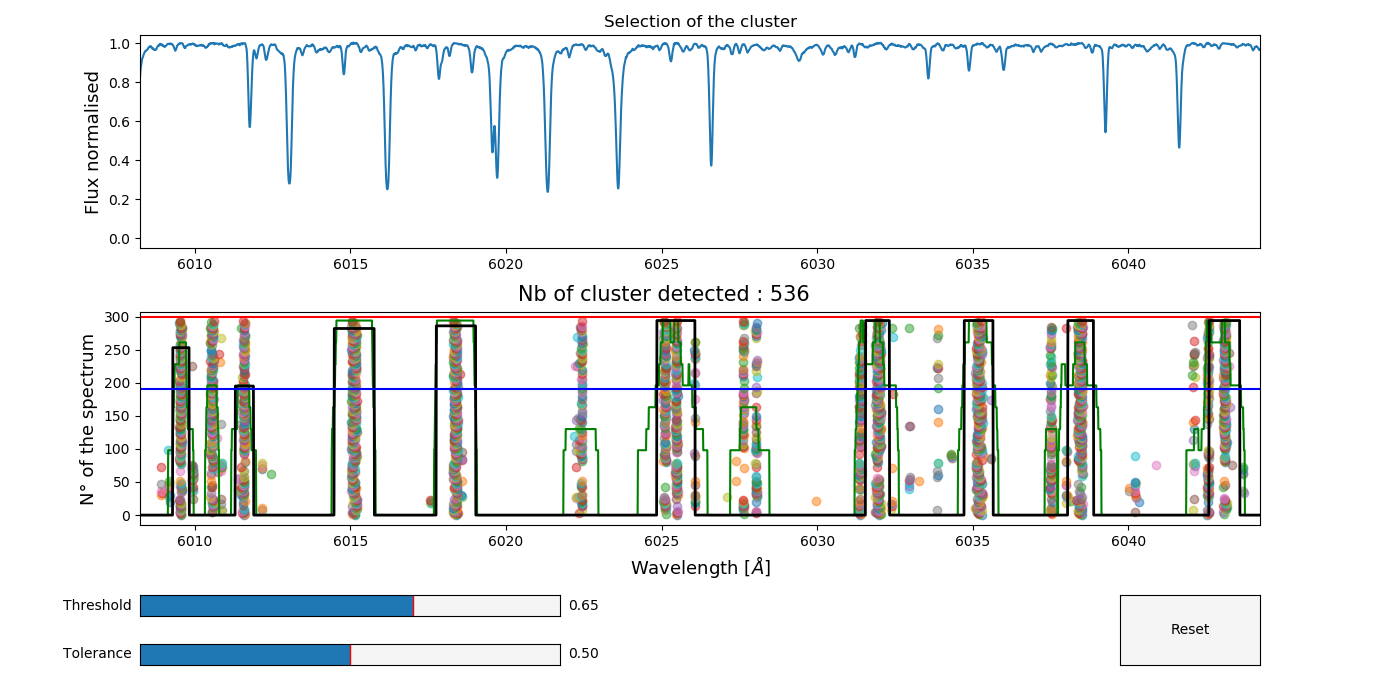}
        \caption{First optional function of RASSINE (\emph{intersect\_all\_continuum}) for reducing spectral time series and stabilising the derived continuum. The anchor points of all the spectra are plotted (colour dots). The density is represented by the green curve. The detected clusters are those with a density higher than the threshold (horizontal blue line) and are represented by the black boxes. This step rejects spurious anchor points such as selected around 6040 $\ang$.}
        \label{FigInt7}
\end{figure*}

\begin{figure*}[h]
\centering
        \includegraphics[width=18.5cm]{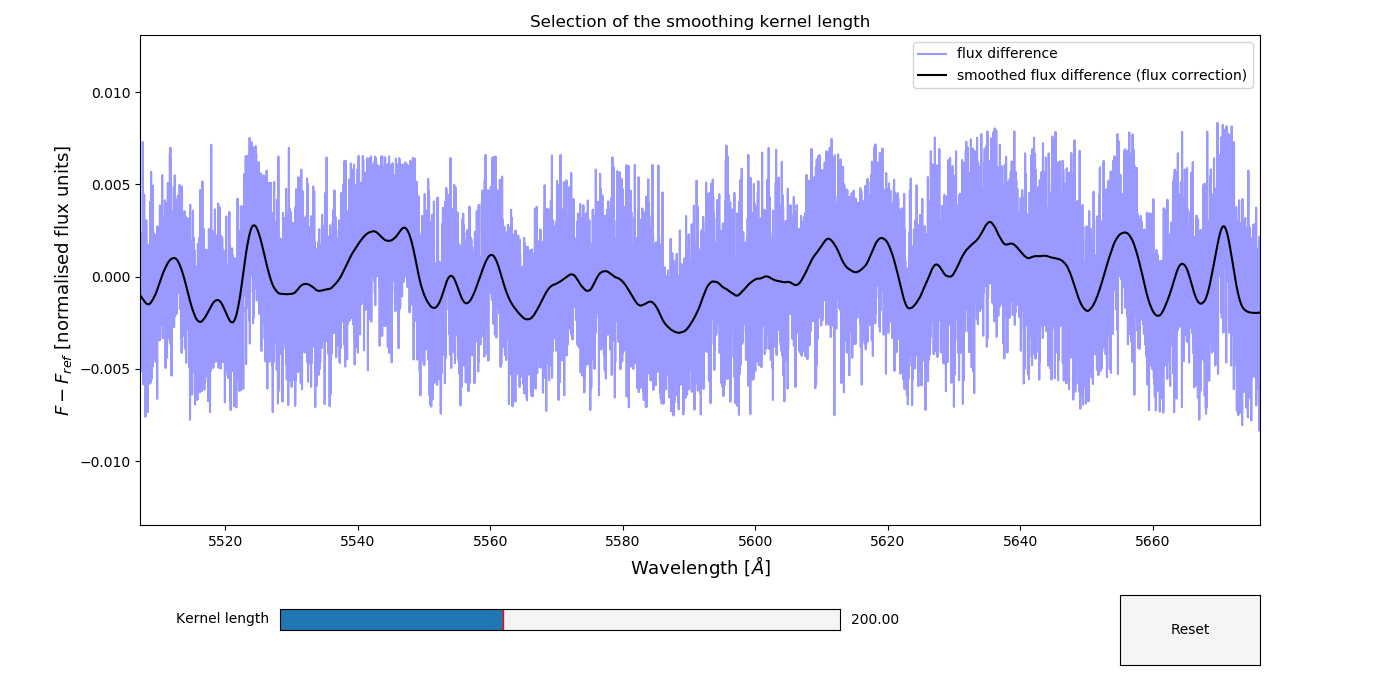}
        \caption{Second optional function of RASSINE (\emph{matching\_diff\_continuum}) for reducing spectral time series and stabilising the derived continuum by applying a Savitzky–Golay filter to the spectral difference with a reference spectrum.}
        \label{FigInt8}
\end{figure*}

In this section, we present the different GUIs that are displayed when graphical feedbacks are activated (\pfeedback{} set to True). The first GUI (Fig.~\ref{FigInt1}) allows smoothing the spectrum, where a slider is used to select the kernel width, and buttons can select the kernel itself. The second GUI (Fig.~\ref{FigInt2}) is used to select the penalty law, which means the minimum and maximum radius of the alpha shape, as well as the penalty law itself. After this step, the code displays the edges of the spectrum (Fig.~\ref{FigInt3}), and the user selects the number of times to flatten the edges. Outliers based on the derivative criterion are then flagged and proposed to be visually inspected by the user (Fig.~\ref{FigInt4}). The user can also manually select (see Fig.~\ref{FigInt5}) the local maxima that they wish to retain or reject by clicking on them. The final product is presented in Fig.~\ref{FigInt6}. For spectral time series, two more GUIs are available. The first optional function (Fig.~\ref{FigInt7}) is launched with the \textit{intersect\_all\_continuum} function. By adjusting the two sliders, the user is shown where the clusters are detected. The second GUI (Fig.~\ref{FigInt8}) is used to select the length of the window used in the Savitzky–Golay filter performed on the spectra difference when the \textit{matching\_diff\_continuum} function is called.

\section{Collections of the RASSINE reduction in automatic mode}

\label{app:collecitons}

\begin{figure*}[h]
\centering
        \includegraphics[width=15.8cm]{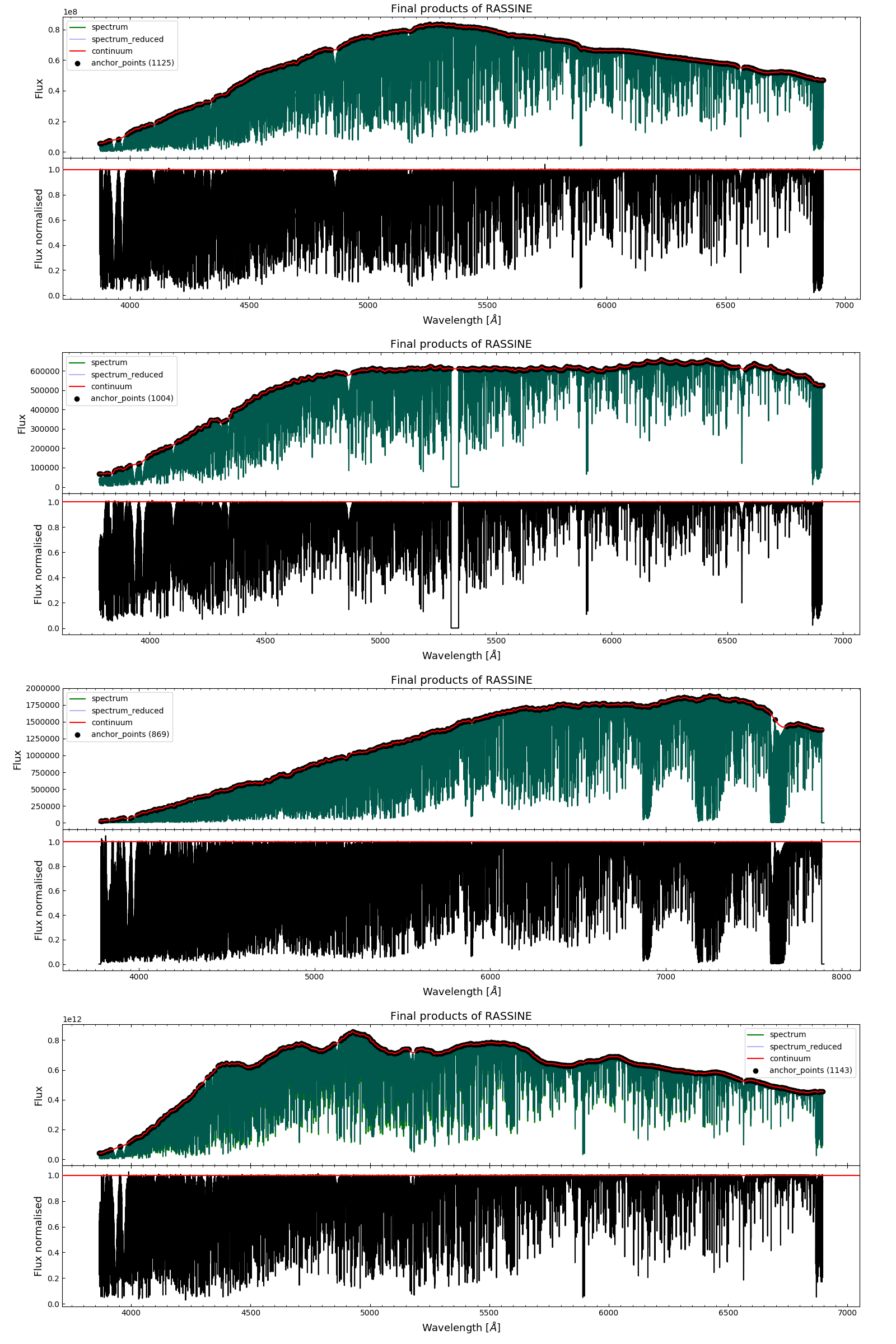}
        \caption{Collection of spectra normalised by RASSINE in automatic mode, $\nu$ = 1, intermediate tension (\pstret{} $=$ 'auto\_0.5'), and Savitzky-Golay filtering with \psbox{} $=6$. Each spectrum was obtained with a different instrument. From top to bottom: HARPS-N, HARPS, ESPRESSO, and CORALIE.}
        \label{FigAllSpec}
\end{figure*}

We already tested RASSINE in automatic mode on spectra obtained with four different instruments that present various shapes of instrumental response. In all the cases, as presented in Fig.~\ref{FigAllSpec}, the continua produced by RASSINE were visually satisfying. All of them were obtained without fine-tuning any parameters except one: the penalty law through the \preg{} parameter. RASSINE therefore handled a smoothed spectral continuum with HARPS-N (first row), an imperfect \emph{\textup{1D}} spectrum reconstruction and CCD gap with HARPS (second row), a telluric forest with ESPRESSO (third row), and an exotic continuum shape with CORALIE (last row). 

\end{appendix}

\end{document}